\documentclass[aip,jcp]{revtex4}
\usepackage{graphicx}
\usepackage{tabularx}
\usepackage{color}
\usepackage{amsmath}
\usepackage{mathtools}
\usepackage{comment}
\newcommand{\be}{\begin{equation}}
\newcommand{\ee}{\end{equation}}
\newcommand{\bea}{\begin{eqnarray}}
\newcommand{\eea}{\end{eqnarray}}
\usepackage{dcolumn}
\usepackage{hyperref}
\usepackage{bm}
\usepackage{epsf}
\usepackage{subfigure}
\usepackage{epstopdf}
\usepackage{amsfonts}
\usepackage{braket}
\usepackage{cancel}

\newcommand{\intinf}{\int_{-\infty}^{\infty}}

\begin{document}

\title{Quantum energy exchange and refrigeration: A full-counting statistics approach} 

\author{Hava Meira Friedman$^1$, Bijay Kumar Agarwalla$^2$, Dvira Segal$^1$}

\address{                    
$^1$Chemical Physics Theory Group, Department of Chemistry, University of Toronto, 80 Saint George St., Toronto, Ontario, Canada M5S 3H6\\
}
\address{
$^2$Department of Physics, Indian Institute of Science Education and Research (IISER)-Pune, Dr Homi Bhabha Rd., Pune, India, 411008} 

\date{\today}

\begin{abstract}
We formulate a full-counting statistics description to study energy exchange in multi-terminal junctions.
Our approach applies to quantum systems that are coupled either
additively or non-additively (cooperatively) to multiple reservoirs.
We derive a Markovian Redfield-type equation for the counting-field dependent reduced density operator.
Under the secular approximation, we confirm that the cumulant generating function satisfies the heat exchange fluctuation theorem. Our treatment
thus respects the second law of thermodynamics.  We exemplify our formalism on a multi-terminal two-level quantum system, and 
apply it to realize the smallest
quantum absorption refrigerator, operating through engineered reservoirs, and achievable only through a cooperative bath interaction model.
\end{abstract}

\maketitle

\section{Introduction}
\label{sec-intro}

The large-scale heat engine was instrumental in the development of classical thermodynamics in the 19th century.
In order to establish the theory of thermodynamics from quantum principles, studies of the nanoscale analogue,
the quantum heat engine (QHE),  are ongoing \cite{Pekola, review-A,reviewK18}.
What is ``quantum" in quantum heat engines \cite{Millen,kur18}? 
The construction of the QHE differs it from the classical heat engine
by having a quantum system, comprising a set of discrete-quantized states,
as the working fluid analogue in many models of such machines \cite{Alan}.
Also, the degrees of freedom of the reservoirs are described by quantum statistical distributions such as 
the Bose-Einstein or Fermi-Dirac functions.
The quantum system is either periodically driven by a classical field,
as in the four-stroke Otto engine \cite{otto1,otto2,otto3},
or operated continuously, to realize e.g. an  absorption refrigerator \cite{QAR}.
It is necessary to consider the ramifications of non-trivial quantum effects
such as quantum coherences  \cite{scully0, scully1, scully2,Raam}, quantum correlations \cite{Huber}
and reservoir engineering e.g. by squeezing \cite{LutzPRL,LutzEPL,bijay-squeeze,Niend1,Niend2},
on the performance of heat engines to potentially overcome classical bounds.

A central aspect of nanoscale quantum heat engines is that they
are not restricted to the weak system-bath coupling limit.
Classical-macroscopic thermodynamics is a weak coupling theory; the effect of the
boundary between the system and its environment is small relative to the bulk behavior.
In contrast, small systems may strongly couple to their environment,
in the sense that the interaction energy between the system and the bath
is comparable to the frequencies of the isolated system.
While quantum thermodynamical machines were traditionally analyzed
under a strict weak-coupling assumption, it is now
recognized that
to properly characterize the performance of quantum engines,
one must develop methods that are not limited in this respect
\cite{DavidS, Kosloff,Cao,Gernot,Gernot2,Eisert-strong,Jarzynski,esposito17,Nazir,Celardo,schmidt,carrega2}. 

What might be the impact of a strong system-bath interaction energy?
Consider, for example, a heat conducting two-terminal nanojunction \cite{reviewSA}.
A weak-coupling (Born-Markov) treatment of the thermal current  
colossally fails beyond the strict weak-coupling regime \cite{nazim}:
While a weak-coupling theory predicts a linear enhancement of the thermal current
with the increase of the system-bath interaction energy \cite{segal-nitzan,segal-master,claire},
a strong-coupling treatment administers a turnover behavior \cite{segal-nitzan,segal-master,segal-nicolin,Ren1,Ren2}.
This crossover, between first enhancement of the current and
subsequent suppression with increasing interaction strength,
appears once the system-bath interaction energy becomes comparable to the system's natural frequencies.

Physically, strong system-bath interactions are responsible for
three-phonon (and more) scattering contributions to thermal transport problems, beyond the weak-coupling resonant term.
Alternatively, rather than focusing on the distinction between strong and weak coupling, one may classify system-bath interaction
operators based on whether they are additive or non-additive in the different reservoirs.
These two classes of models, additive and non-additive, realize distinctive energy transport characteristics and refrigeration 
as we show in this work.

An autonomous absorption refrigerator transfers thermal energy from a cold bath to a hot bath without an input power,
by using thermal energy provided from a so-called work reservoir.
In a recent study \cite{Anqi}, we demonstrated that the smallest system, 
a qubit, is incapable of operating as a quantum absorption refrigerator (QAR) when the baths are coupled weakly-additively to the system.
However, we demonstrated that by coupling the system in a non-additive manner to three heat baths, 
which are spectrally structured, a cooling function was achieved. 
Moreover, we showed that the system reached the Carnot bound
when the reservoirs were characterized by a single frequency component.
This study thus clearly illustrates that non-additive models can bring in a new function, which is missing in additive scenarios.
On the other hand, as was pointed out in Ref. \cite{Anqi} and  illustrated earlier on in Refs.
 \cite{segal-nitzan,segal-master,segal-nicolin}, the dynamics of non-additive models can be treated with 
standard kinetic quantum master equations,  albeit with a non-additive, baths-cooperative dissipator.

The objective of this paper is to present a rigorous, thermodynamically consistent formalism for the
calculation of energy exchange (current and cumulants) in additive and non-additive interaction models, thus
develop the groundwork of the qubit-QAR presented in Ref. \cite{Anqi}.
%
Our formalism utilizes a full-counting statistics (FCS) approach that
provides cumulants of energy exchange to all orders \cite{esposito-review,hanggi-review,bijay-wang-review,carrega,carrega2}. 
In this method, rather than focusing directly on the averaged heat current,
the first cumulant,  we analyze the probability distribution of exchanged energy with each bath,
within a certain interval of time $t$.
More specifically, we study the Fourier transform of this probability distribution function,
the so-called characteristic function.  
This function can be derived from an equation of motion for
the counting-field dependent reduced density operator,
which we organize in the form of a Redfield equation.
The formalism is applied to treat both additive and non-additive interaction models. 
Under the secular approximation, our derivation is thermodynamically consistent: We confirm that the derived
cumulant generating function (CGF) satisfies the entropy production fluctuation theorem,
which is a microscopic statement of the second law  \cite{esposito-review,hanggi-review}.
We exemplify our work on the two-state system, culminating this study in the exploration of the cooling 
performance of a qubit refrigerator.

The paper is organized as follows. We motivate the study of non-additive interaction models in Sec. \ref{sec-NADD}.
Sec. \ref{sec-method} contains our method development.
We define the model, derive the counting-field dependent Redfield-type equation
and use it to calculate the CGF.
Closed-form expressions for the energy current in two-terminal setups,
assuming either additive or non-additive system-bath interaction operators are also presented.
In Sec. \ref{sec-TLS}, we exemplify our analytical results on a two-state system and present simulation results of the two-terminal 
nonequilibrium spin-boson model. The application of our formalism to study a quantum absorption refrigerator is
given in Sec. \ref{sec-engine}. 
We summarize our work in Sec. \ref{sec-summary}. 
Throughout the paper we work with units where $\hbar=1$, $k_B=1$.

\begin{figure}
\includegraphics[width=15cm]{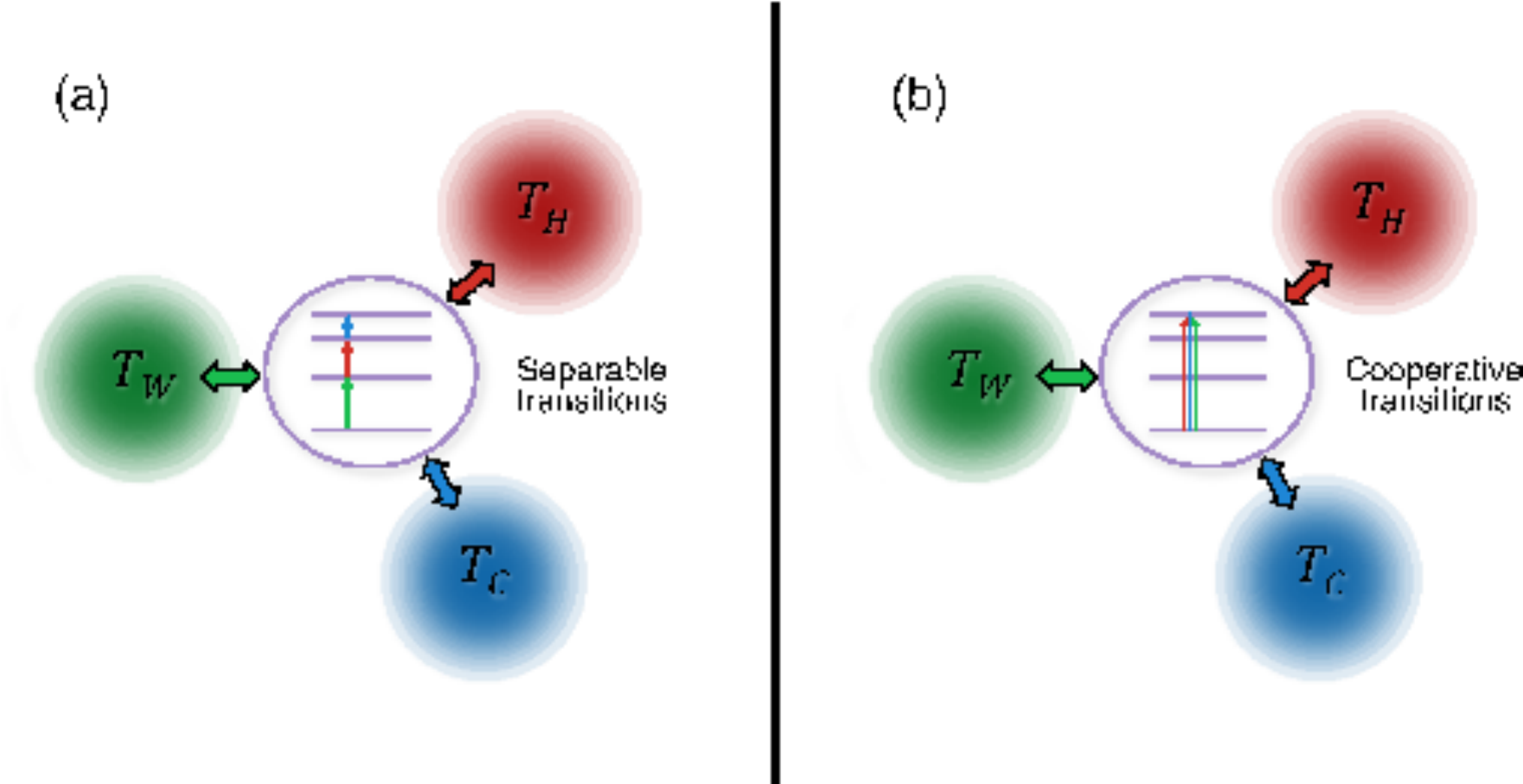} 
\vspace{-4mm}
\caption{
Examples of model systems that can be treated by our approach.
(a) The system-bath coupling operator is additive in the individual baths, leading to sequential transitions
that are dictated separately by each bath.
(b) The system-bath interaction is non-additive in the individual baths, resulting in convoluted, cooperative transition rate constants.
}
\label{scheme}
\end{figure}

\section{Additive and non-additive Interaction Models}
\label{sec-NADD}

In standard models of open quantum systems, under the weak-coupling approximation
the different baths dictate additive decoherence and dissipative dynamics  \cite{Breuer, Nitzan}.
As a result, heat currents flow independently between the quantum system and 
each individual bath \cite{Alicki,segal-nitzan,segal-master}.
Let us clarify this point. The commonly used open-quantum-system Hamiltonian,
with a system $\hat H_S$ coupled to $N$ reservoirs is
\bea
\hat H  = \hat H_S
        + \sum_{\nu=1}^N \hat H_\nu
        +   \hat H_{int} ,
        \,\,\,\
\hat H_{int}=\sum_{\nu=1}^N \hat S_{\nu} \otimes \hat V_{\nu} \,.
\label{eq:H0}
\eea
Here,  $\hat S_{\nu}$ is a system operator that is coupled to the $\nu$ bath's degrees of freedom
and $\hat V_{\nu}$ is an operator of the $\nu$ bath.
Note that there is an inherent assumption of additivity of the interaction of different reservoirs with the system.
Assuming weak system-bath coupling,
an equation of motion can be derived for the reduced density matrix of the system, $\sigma$,
valid to second order in the interaction Hamiltonian.
Under the Markovian approximation, we receive the quantum master equation (QME) ($\hbar\equiv 1$)
\bea
\frac{d\sigma}{dt}=-i[\hat H_S,\sigma] + \sum_{\nu=1}^N D_{\nu}[\sigma(t)],
\label{eq:dyn}
\eea
with the dissipators $D_{\nu}[\sigma(t)]$ that contain information about the $\nu$th bath's effect on the system.
Decoherence and relaxation dynamics is therefore additive in the different baths, i.e., the total relaxation rate
of the system is the sum of relaxation rates induced by each bath.
From Eq. (\ref{eq:dyn}), the rate of energy relaxation from the system can be written as
\bea
\langle \dot {\hat {H}}_S \rangle =  \frac{d}{dt} {\rm Tr}\left[{\hat H_S \sigma(t)}\right] = \sum_{\nu=1}^N  {\rm Tr}\left[\hat H_S D_{\nu}[\sigma(t)]\right].
\label{eq:Hsdot}
\eea
We can readily identify the currents flowing towards the system from the different baths
as $\langle J_{\nu}(t)\rangle\equiv{\rm Tr}[\hat H_S D_{\nu}[\sigma(t)]]$.
This analysis is obviously limited to the additive interaction Hamiltonian---and when the dynamics can be recast into (\ref{eq:dyn}).

In this manuscript we present a thermodynamically consistent formalism for the
calculation of energy exchange  in situations that do not necessarily follow the evolution equation (\ref{eq:dyn}).
In particular, we  consider two classes of system-bath interaction models,
shown as schemes in Fig. (\ref{scheme}),
$\hat H_{int}=\gamma \sum_p \hat S^p \otimes \hat V^p_N$,
with either $\hat V^p_N=\sum_{\nu=1}^N\hat B^p_{\nu}$ (panel a) or $\hat V^p_N=\Pi_{\nu=1 }^N\hat B^p_{\nu}$ (panel b).
Here, $\hat S^p$ is a system operator,  $\hat B^p_{\nu}$ is a $\nu$-bath operator,
and $\gamma$ is introduced to keep track of the perturbative expansion.
The baths may couple to multiple system's operators, counted by the index $p$.
The first model is additive (ADD) and separable in the reservoirs' interaction operators $\hat B_{\nu}^p$, as in Eq. (\ref{eq:H0}).
The second model is non-additive (NADD),  or multiplicative, in $\hat B^p_{\nu}$ allowing for a cooperative effect of the baths.
In both cases, however, we can study the system's dynamics and its energy transport characteristics by
using a perturbation theory treatment up to second order in $\gamma$ and calculating the CGF.

We distinguish here between additive and non-additive system-bath interaction models
while one usually makes a division
between weak- and strong-coupling limits---for additive Hamiltonians of the form (\ref{eq:H0}).
In the context of quantum transport,
additive models were examined in the weak-coupling approximation,
then improved by including higher order interaction effects,
e.g. by exercising perturbation theory to the fourth order \cite{wang4},
using dressing  \cite{segal-nitzan, segal-master, segal-nicolin} or mapping approaches \cite{gernot,nazir}.
Recently developed numerically-exact simulation methods \cite{thoss,segal-infpi,saito,tanimura,brandes,schmidt}  interrogate strong coupling effects,
though they often have limitations in system size.
Here, we analyze a general interaction model for multiple baths, and use it
to study additive and non-additive interaction models.
In both cases, we demonstrate how to derive the energy currents and noise flowing out of each bath, and
our analytical results can provide physical insights to the underlying mechanisms for energy exchange.

Intuitively,  the ADD interaction model is applicable when the reservoirs are physically separated from each other
and therefore do not interact.  The NADD model can show up in different scenarios.
For example, in the treatment of the nonequilibrium (two-bath) spin-boson model
one can perform a unitary (polaron) transformation of the Hamiltonian resulting 
in a product of displacement operators for the system-bath coupling, then proceed to
study the dynamics under the Born-Markov approximation \cite{segal-nitzan, segal-master,segal-nicolin}.
The bare system-bath interaction energy is incorporated inside the displacement operator,
which is an exponential function, thus its effect is included to high orders.
NADD models can also be accomplished by engineering many-body Hamiltonians based on e.g. resonant conditions
and selection rules \cite{QAR,plenio}. In Ref. \cite{Alicki}, the NADD Hamiltonian
represents a chemical engine where reactants are destroyed and a chemical product is created,
along with an excitation.

The NADD model may realize behaviors that cannot be captured by an ADD  Hamiltonian.
We exemplify this fact by studying a two-level system to realize the smallest quantum absorption refrigerator,
with a qubit as its working fluid analogue \cite{Anqi}.
A cooperative bath behavior is what allows for refrigeration.

\section{Method: full-counting statistics for energy exchange}
\label{sec-method}

We present a projection operator formalism \cite{nakajima,zwanzig} 
for deriving the characteristic function for quantum energy exchange
between a system and multiple attached thermal reservoirs. 
In steady state, we ultimately obtain the cumulant generating function 
for quantum energy exchange. This function provides the energy current 
and any higher-order cumulant of the current such as the noise power.

Assuming weak system-bath coupling and memoryless reservoirs, 
the procedure is formulated in the language of a counting-field dependent 
Markovian quantum master equation---the Redfield equation---generalized to describe the
full-counting statistics of energy exchange in a multi-terminal geometry.
Our formalism is valid in the weak coupling limit 
but allows for system transitions induced cooperatively by $N$ baths. 

\subsection{Hamiltonian}
\label{subsec-H}

Our discussion begins with a general open quantum system Hamiltonian, 
a system $\hat H_S$ coupled to $N$ reservoirs, 
\bea
\hat H	= \hat H_S
+ \sum_{\nu=1}^N \hat H_\nu  
+ \gamma \sum_p \hat S^p \otimes \hat V^p_N.
\label{eq:H}
\eea
Here,  $\hat S^p$ is a system operator
and $\hat V^p_N$ is an operator of the environment (bath) that depends on the nature and number of reservoirs that 
are included in the model. Note that the summation over $p$ allows for many possibilities 
for system coupling to the reservoirs in different configurations
with operators $\hat S^p$. 
The Hamiltonian is time-independent.
Therefore, in the steady-state limit energy exchange between the system and the reservoirs corresponds to heat  transfer.

To study energy transfer over the time interval $[0,\,t]$, between the system and the different reservoirs, 
we write down the energy current from reservoir $\nu$ as a change in the $\nu$ bath energy $J_{\nu}(t)=-\frac{d\hat H_{\nu}(t)}{dt}$.
Operators are written in the Heisenberg representation, $\hat A(t) = \hat U^{\dagger}(t) \hat A \hat U (t)$,
where $\hat U(t) = e^{-i\hat Ht}$. Therefore, the total energy exchange at the $\nu$ terminal is given by the integrated current
\bea
Q_\nu(t,t_0=0) = 
\int _0^t J_{\nu}(t')dt' =        \hat H_\nu(0) -  \hat H_\nu(t).
\label{eq:Q}
\eea
Employing the two-time measurement protocol \cite{esposito-review,bijay-wang-review}, 
the characteristic function for energy transfer at multiple baths takes the form
\bea
{\cal Z}(\lambda=\{\lambda_\nu\},t) \equiv 
{\rm Tr} \left[ e^{i\sum_\nu \lambda_\nu \hat H_\nu(0)} e^{-i\sum_\nu\lambda_\nu \hat H_\nu(t)} \rho(0) \right],
\label{eq:Z1}
\eea
where we have introduced $\lambda_\nu$ as a real-valued parameter for counting energy
at the $\nu$ reservoir, 
and $\rho(0)$ is the total density matrix at the initial time $t=0$. 
In the long-time limit, the CGF is defined as 
\bea
{G}(\lambda) \equiv \lim_{t \to \infty} \frac{1}{t} \ln {\cal Z} (\lambda).
\label{eq:CGF1}
\eea
Derivatives of $G(\lambda)$ with respect to $(i\lambda)$ yield the steady-state energy current cumulants.
We now recast Eq. (\ref{eq:Z1}) in the structure of the Liouville equation
by explicitly writing down the time evolution operators, factorizing the exponentials, 
using the cyclic property of the trace operation  
and the assumption that $\rho(0)$ commutes with $\hat H_\nu$. 
We compactly write
\bea
{\cal Z}(\{\lambda_\nu\},t) 
&=& {\rm Tr} \left[ 
\hat U^{-\lambda}(t) \rho(0)  
\hat U^{\lambda\,\dagger}(t)\right]  
\nonumber\\
&=&
{\rm Tr} \left[ \rho^{\lambda}(t)
\right].
\label{eq:Zcompact}
\eea
Here, we define the time evolution operators 
$\hat U^{\lambda\,\dagger}(t) \equiv e^{i\sum_\nu \lambda_\nu \hat H_\nu/2} \hat U^{\dagger}(t) e^{-i \sum_\nu \lambda_\nu \hat H_\nu/2}$, 
and $\hat U^{-\lambda}(t) \equiv e^{-i\sum_\nu \lambda_\nu \hat H_\nu/2} \hat U(t) e^{i \sum_\nu \lambda_\nu \hat H_\nu/2}$. 
In our convention, the sign of the subscript determines the sign of the counting terms in the exponent.
We also define the counting-field dependent time evolution operator,
\bea
\hat U^{-\lambda}(t) \equiv e^{-i \hat H^{-\lambda} t},
\label{eq:U_xi}
\eea
with 
$\hat H^{-\lambda} \equiv e^{-i\sum_\nu \lambda_\nu \hat H_\nu/2} \hat H e^{i\sum_\nu \lambda_\nu \hat H_\nu/2}$.
Since $\hat H_\nu$ commutes with all operators on the right hand side of Eq. (\ref{eq:H}), except $\hat V_N^{p}$, 
we obtain a counting-dependent total Hamiltonian as 
\bea
\hat H^{-\lambda}	
= \hat H_S + \sum_{\nu=1}^N \hat H_\nu  
+ \gamma \sum_p \hat S^p \otimes \hat V_N^{p,-\lambda},
\eea
where $\hat V_N^{p,-\lambda} = e^{-i\sum_\nu \lambda_\nu \hat H_\nu/2} \, \hat V_N^p \, e^{i\sum_\nu \lambda_\nu \hat H_\nu/2}$. 
In order to ultimately reach the CGF,  we describe the counting-field dependent Nakajima-Zwanzig equation in Appendix A, and
present in Sec. \ref{subsec-QME} a second order quantum master equation to obtain the counting-field dependent density matrix in Eq. (\ref{eq:Zcompact}). 

\subsection{Counting-field dependent second-order quantum master equation}
\label{subsec-QME}

Continuing from Eq. (\ref{eq:Zcompact}), we switch to the interaction representation and define the counting-field density matrix, 
\bea
\rho^\lambda_I(t) \equiv 
\hat U_I^{-\lambda}(t) \rho(0) \hat U_I^{\lambda\,\dagger}(t),
\label{eq:rhoI}
\eea
with $\hat U_I^{-\lambda}(t) = \mathrm T\, e^{-i \int_0^t \hat H_{int}^{-\lambda}(\tau) d\tau}$, 
$\hat H_{int}^{-\lambda}(\tau)=\gamma \sum_p \hat S^p(\tau) \otimes \hat V_N^{p,-\lambda}(\tau)$. 
Operators are now written in the interaction representation,
$\hat A(t)=\hat U_0^\dagger(t)\hat A \hat U_0(t)$ where $\hat U_0(t) = e^{-i \hat H_0 t}$, $\hat H_0=\hat H_S + \sum_{\nu=1}^N \hat H_\nu$. 
Eq. (\ref{eq:rhoI}) can be written as a differential equation,
\bea
\dot \rho^\lambda_I (t) = 
i\,  \rho^\lambda_I(t) \, \hat H^{\lambda}_{int}(t)  - i \, \hat H^{-\lambda}_{int}(t) \, \rho^\lambda_I(t), 
\label{eq:rho_com}
\eea
which reduces to the quantum Liouville equation when $\lambda=0$.
For convenience, we omit below the interaction representation descriptor `$I$' moving forward. 

We follow the standard derivation of the second order Markovian master equation \cite{Breuer}---while 
taking into account the counting information. 
An important assumption is that  the thermal baths are prepared in a canonical thermal state. For details, see Appendix A.
The result of this treatment is the counting-field dependent Redfield-type equation \cite{Nitzan} for
$\sigma^\lambda(t) \equiv  {\rm Tr}_{\textrm{B}} \left[ \rho^\lambda (t)\right]$, satisfying (Schr\"odinger representation),
\bea
\dot \sigma_{nm}^\lambda (t) = -i[\hat H_S,\sigma(t)]_{nm} +
	 \sum_{p,q,j,k} \Big[ 
&-& \sigma_{km}^\lambda(t) R^{p,q^{(+)}}_{N;\,{nj,jk}}(E_{k,j}) 
 -  \sigma_{nj}^\lambda(t) R^{p,q^{(-)}}_{N;\,{jk,km}}(E_{j,k}) 
\nonumber \\
&+& \sigma^\lambda_{jk}(t)  \left( R^{q,\lambda,p,-\lambda^{(-)}}_{N;\,{km,nj}}(E_{k,m})
 + R^{q,\lambda,p,-\lambda^{(+)}}_{N;\,{km,nj}}(E_{j,n}) \right)\Big].
\label{eq:Redfield_counting}
\eea
Here, $E_{m,n} = E_m - E_n$ and $E_n$ are the energy eigenstates of the system Hamiltonian $\hat H_S$. 
$p$ and $q$ sum over the different operators that couple to the system. 
The other indices, $j,k$, count eigenstates of the system. The different terms in this equation are defined in Appendix A.

Eq. (\ref{eq:Redfield_counting}) can be further simplified by performing the secular approximation, 
so as to decouple population and coherence dynamics.
The population dynamics, $p_{n}(t) \equiv \sigma_{nn}(t)$, then follows
\bea
\dot p_{n}^\lambda (t) = 
		-  p_{n}^\lambda(t) \sum_{p,q,j}  M^{p;q}_{N;\,{nj,jn}} (E_{n,j})  
		+  \sum_{p,q,j} p^\lambda_{j}(t) \, M^{q,\lambda;p,-\lambda}_{N;\,{nj,jn}} (E_{j,n}),
\label{eq:population_under_SA_Rs}
\eea 
with
\bea
M^{p,\lambda;p^\prime, \lambda^\prime}_{N; \, {ab,cd}}(s)
\equiv
\gamma^2 \, S^p_{{ab}} S^{p^\prime}_{{cd}} \, \langle \hat V_{N}^{p,\lambda}(s) \hat V_{N}^{p^\prime,\lambda^\prime}(0) \rangle.
\label{eq:Ms}
\eea
The averages are performed with respect to the initial condition.
The rate constants are Fourier transforms of the correlation functions,
$M^{p,\lambda;p^\prime,\lambda^\prime}_{N;\,{ab,cd}}(\omega) 
\equiv 
\intinf e^{i \omega s} \, M^{p,\lambda;p^\prime,\lambda^\prime}_{N;\,{ab,cd}}(s) \, ds
$. 
Eq. (\ref{eq:population_under_SA_Rs}) is a Markovian-secular counting-field dependent quantum master equation 
for the system population dynamics, with the system 
coupled to $N$ baths.
We can write it down in a compact matrix form with the counting-dependent Liouvillian $\mathcal{L}^\lambda$ as  
\bea
\ket{\dot p^\lambda (t)} = \mathcal{L}^\lambda \ket{p^\lambda(t)}.
\label{eq:matrix_QME}
\eea
Analogous derivations of the counting-field dependent reduced density operator,
for charge transfer problems, were performed in e.g.
Refs. \cite{SF-segal,bijay-segal-fcs,bijay-QME,hava1,olaya}. 
Recall that Eq. (\ref{eq:matrix_QME}) gives us the characteristic function 
$\cal Z$ from Eq. (\ref{eq:Zcompact})  
and ultimately the CGF via Eq. (\ref{eq:CGF1}). 

While our derivation is standard, our objective here has been to highlight that 
Eqs. (\ref{eq:Redfield_counting}) and (\ref{eq:population_under_SA_Rs}) are 
valid for both ADD and NADD interaction Hamiltonians.
These two equations are central to this paper and can be used to calculate 
energy transfer cumulants for a general open quantum system setup---given by the Hamiltonian in Eq. (\ref{eq:H}). 

\subsection{Energy current}
\label{subsec-Current}

We organize a closed expression for the energy current from Eq. (\ref{eq:matrix_QME})
by differentiating $G$ with respect to the counting-field $\lambda$. 
The characteristic function is given by a trace over the system states. 
From Eq. (\ref{eq:Zcompact}),  $\mathcal Z(\lambda,t)=\langle 1| p^{\lambda}(t)\rangle$
with $\langle 1|=(1,1,...1)^T$ as the identity vector.
As mentioned before, differentiating $G(\lambda)$ with respect to $(i\lambda_\nu)$ returns the steady-state current between the system and the $\nu$ bath, $\langle J_{\nu}\rangle=
\lim_{t \to \infty} \frac{1}{t} \frac{\partial \ln {\cal Z} (\lambda,t)}{\partial (i\lambda_{\nu})}|_{\lambda=0}$,
\bea
\langle J_{\nu}\rangle
=
\langle 1 | \frac{\partial \mathcal{L}^{\lambda }}{\partial (i\lambda_{\nu})}
\Big|_{\lambda=0} |p^{ss}\rangle
=
\sum_{n,j} \left( \frac{\partial \mathcal{L}^{\lambda }}{\partial (i\lambda_{\nu})} \Big|_{\lambda=0} \right)_{nj}p_j^{ss},
\label{eq:Jpop}
\eea
with the steady state populations $p_j^{ss}$, which we get by solving Eq. (\ref{eq:matrix_QME}) in the long-time limit. 
The full-counting statistics approach thus provides a rigorous working expression of the current for ADD and NADD models, at the same footing. 
Higher order cumulants can also be calculated by taking higher order derivatives of $G$, 
such as the second cumulant,  the noise power, 
calculated in Sec. \ref{subsec-TLS-current-noise}. 
Beyond the Markov approximation,
it is useful to note that non-Markovian effects do not enter the steady state current, but they do 
lead to corrections to higher order cumulants. Such non-Markovian effects can be evaluated with techniques
developed for the study of full counting statistics in charge transport \cite{Fazio}.

So far, we considered a system coupled to $N$ thermal baths,  with $N$ counting parameters, $\lambda_1$, $\lambda_2$, ..., $\lambda_\nu$, ..., $\lambda_N$. 
For simplicity,  in what follows we count energy at a single bath only, 
the $\nu$ bath, and denote  $\lambda =  \lambda_\nu$. 
We study two different types of system-bath couplings; 
additive $\hat V_N = \sum_{\nu=1}^N \hat B_\nu$ 
and non-additive $\hat V_N = \prod_{\nu=1}^N \hat B_\nu$.
The latter case may be realized by building up a compound interaction operator, e.g., through a unitary transformation 
of operators. 
Further, for simplicity, we treat a single system-bath operator, i.e., we ignore
the $p,q$ summation in  Eq. (\ref{eq:population_under_SA_Rs}). 
Finally, we henceforth ignore the $\gamma^2$ factor. 

The counting-dependent Liouvillian $\cal L^\lambda$ is made of correlation functions for the $\nu$ bath
(\ref{eq:Ms}),
\bea
M_{\nu; \, {nj,jn}}^{\lambda;-\lambda}(s) 
=\langle e^{i\hat H_\nu\lambda} e^{i\hat H_\nu s}\hat B_\nu e^{-i\hat H_\nu s}e^{-i\hat H_\nu \lambda} \hat B_\nu \rangle S_{nj}S_{jn}.
\eea
Its Fourier transform in frequency domain gives the relation,
\bea
M_{\nu; \, {nj,jn}}^{\lambda;-\lambda}(\omega) =  
e^{-i \omega \lambda} M_{\nu;\,{nj,jn}}(\omega),
\label{eq:MLambdaReln}
\eea
which is useful for calculating  derivatives in Eq. (\ref{eq:Jpop}). 
It is also useful to note the detailed balance relation with counting parameters for 
a specific bath $\nu$ at inverse temperature $\beta_\nu$, 
\bea
	M_{\nu;\,nj,jn}^{\lambda;-\lambda}(\omega) = e^{(\beta_\nu-i\lambda)\omega} M_{\nu;\,nj,jn}(-\omega).
\label{detailed_balance}
\eea
For details, see Appendix B.

\subsubsection{Additive bath interaction model: $\hat V_N = \sum_{\nu=1}^N \hat B_\nu$}
\label{subsubsec-method-add}

In the ADD case,  $\hat H_{int} ^{\lambda}= \gamma \, \hat S \otimes \hat V_N^\lambda$ 
with $\hat V_N^\lambda(t) = \hat B_1(t) + \hat B_2(t)+ ... + \hat B_\nu^\lambda(t)+ ... + \hat B_N(t)$. 
Recall that baths' operators are written in the interaction representation, 
the baths are initially uncorrelated and we assume that $\langle \hat V_N \rangle=0$. 
This results in the rate constants in Eq. (\ref{eq:population_under_SA_Rs}) being additive in the $N$ reservoirs,
\bea
M_{N; \,{nj,jn}}^{\lambda;-\lambda}(E_{j,n})
= M_{1; \,{nj,jn}}(E_{j,n}) +...+ M_{\nu;\,{nj,jn}}^{\lambda;-\lambda}(E_{j,n})+...+M_{N; \,{nj,jn}}(E_{j,n}).
\eea
The Liouvillian is then given by adding up all contributions,
\bea
\ket{\dot p^\lambda (t)} 
=\left(\mathcal{L}_1+...+ \mathcal{L}^{\lambda}_\nu +...   + \mathcal{L}_N\right)    \ket{p^\lambda(t)}.
\label{eq:L2}
\eea
Using Eqs.  (\ref{eq:population_under_SA_Rs}), (\ref{eq:Jpop}) and (\ref{eq:MLambdaReln}) we receive 
for the energy current 
\bea
\langle J_\nu \rangle &=& \sum_{n,j} p_j^{ss} E_{n,j} |S_{nj}|^2 
\intinf ds \, e^{iE_{j,n}s} \langle \hat B_\nu(s) \hat B_\nu(0)\rangle  \nonumber \\
&=& {\rm Tr} [\hat H_S \mathcal{L}_\nu \, p^{ss}],
\label{eq:JE}
\eea
with $\mathcal{L}_\nu \, p^{ss}$ as the dissipator of the dynamics due to the $\nu$ reservoir.
It is worth commenting that this relation can be immediately derived from the energy balance equation for
the system energy operator, $\frac{d {\rm Tr}[ \hat H_S \sigma(t)]} {dt}=\sum_{\nu=1}^N \langle J_\nu\rangle$,
as discussed in Sec. \ref{sec-NADD}.
Nevertheless, the formalism presented here can be used to feasibly provide higher order cumulants.

\subsubsection{Non-additive bath interaction: $\hat V_N = \prod_{\nu=1}^N \hat B_\nu$}
\label{subsubsec-method-nadd}
We now assume that the system is coupled to the environment according to  the following form, 
$\hat V_N^\lambda(t) = \hat B_1(t) \otimes \hat B_2 \otimes ... \otimes \hat B_\nu^\lambda(t) \otimes ... \otimes \hat B_N(t)$. 
This structure arises e.g. in the study of the spin-boson model after a polaron transformation \cite{segal-nicolin}.
We insert this product form into the definition of the correlation function and arrive at
\bea
M^{\lambda;\lambda^\prime}_{N; \,{ab,cd}}(s) 
= \langle \hat B_1(s) \hat B_1(0) \rangle \langle \hat B_2(s) \hat B_2(0) \rangle ... \langle \hat B_\nu^\lambda(s) \hat B_\nu^{\lambda^\prime}(0) \rangle...\langle \hat B_N(s) \hat B_N(0) \rangle \, 
S_{{ab}} S_{{cd}},
\label{eq:Mprod}
\eea
since the bath initial state is factorized, $\rho_{\textrm{B}} = \prod_{\nu=1}^N \rho_\nu$.
In frequency domain, the correlation function turns into a convolution,
\bea
M_{N;\,{nj,jn}}^{\lambda;-\lambda}(E_{j,n})= \frac{1}{(2\pi)^{N-1}}\int_{-\infty}^{\infty} d\omega_2 \,... \,d\omega_\nu \,...\,d\omega_N\, \Big[
M_{1;\,{nj;jn}}( E_{j,n} - \omega_2 - ... - \omega_\nu -...-\omega_N) \, M_{2;\,{nj,jn}}(\omega_2) \nonumber \\
... \, M_{\nu;\,{nj,jn}}^{\lambda;-\lambda}(\omega_\nu)\,...\, M_{N;\,{nj,jn}}(\omega_N) \Big].
\label{eq:rateMN}
\eea
This rate constant describes a cooperative effect: The system makes a transition between the $n$ and $j$ states,
and energy is absorbed or released  simultaneously-partially to the $N$ baths.
From Eq. (\ref{eq:population_under_SA_Rs}) and the relation in (\ref{eq:MLambdaReln}), 
we receive the energy current from Eq. (\ref{eq:Jpop}), 
\bea
\langle J_\nu \rangle= -\frac{1}{(2\pi)^{N-1}}
\sum_{n,j}  p_j^{ss} \int_{-\infty}^{\infty} d\omega_2 \,... \,d\omega_\nu \,...\,d\omega_N\, (\omega_\nu) \,\Big[
M_{1;\,{nj;jn}}( E_{j,n} - \omega_2 - ... - \omega_\nu -...-\omega_N) \, M_{2;\,{nj,jn}}(\omega_2) \nonumber \\
... \, M_{\nu;\,{nj,jn}}(\omega_\nu)\,...\, M_{N;\,{nj,jn}}(\omega_N) \Big].
\label{eq:J_nadd}
\eea
We emphasize that this result holds without specifying the system, bath or the system-bath interaction Hamiltonian, 
aside from it being a product form. 
%

\section{The Nonequilibrium two-level system}
\label{sec-TLS}

\subsection{Cumulant generating function}
\label{subsec-TLS-CGF}

Spin-bath models are central to the theory of open quantum systems \cite{Weiss-book,Leggett-spin}.
In particular, the spin-boson model has found immense applications in condensed phases physics, 
chemical dynamics, quantum optics and quantum technologies \cite{Weiss-book}. It serves
to develop approximation schemes perturbative and non-perturbative in the system-bath coupling strength, see for example recent studies \cite{LeHur,Grifoni}.
Beyond questions over quantum decoherence, dissipation, and thermalization,
which can be addressed within the single-bath spin-boson model,
the two-bath, nonequilibrium spin-boson (NESB) model has been put forward  as a minimal model for exploring the fundamentals
of thermal energy transfer in anharmonic nano-junctions \cite{segal-nitzan}.
When the two reservoirs are maintained at different temperatures away from linear response, nonlinear
functionality such as the diode effect can develop in the junction \cite{segal-nitzan,segal-master, claire, segal-nicolin}.
From the theoretical perspective, the NESB model is a rich platform for developing methodologies in 
nonequilibrium quantum dynamics. For recent comprehensive studies, see e.g. Refs. \cite{nazim,bijay-majorana,saito18}.

In this section, we consider  a nonequilibrium spin-bath model: a qubit coupled via a $\hat \sigma_x$ operator to $N$ reservoirs.
Our objective is to work with the formalism of Sec. \ref{sec-method} and derive expressions for the current and noise in ADD and NADD spin-bath models.
For simplicity, we continue with a single counting parameter $\lambda=\lambda_\nu$, which counts energy at the $\nu$ contact. 
The system Hamiltonian is written in its energy eigenbasis, and the interaction operator with the bath is off-diagonal. We do not specify the reservoirs and their interaction
with the qubit,
\bea
\hat H_S &=& \frac{\omega_0}{2}\hat \sigma_z = \frac{\omega_0}{2} \left( \ket{1}\bra{1} - \ket{0}\bra{0} \right) ,
\nonumber \\ 
\hat H_{int} &=& \hat S \otimes \hat V_N, \,\,\,  \hat S = \hat \sigma_x = \ket{0}\bra{1} + \ket{1}\bra{0}.
\label{H-TLS}
\eea
In the language of the general discussion, $S_{00} = S_{11} = 0$,   $S_{01}=S_{10}=1$, 
and $E_{1,0}=\omega_0$, $E_{0,1}=-\omega_0$.  
The QME (\ref{eq:population_under_SA_Rs})  thus reduces to
\bea
	\begin{pmatrix} \dot p_0^\lambda (t) \\ \dot p_1^\lambda (t) \end{pmatrix} = 
		\begin{pmatrix}
			-M_{N}(-\omega_0)  & 
			M_{N}^{\lambda;-\lambda}(\omega_0) 	\\
 			M_{N}^{\lambda;-\lambda}(-\omega_0) 	&
			-M_{N}(\omega_0) 		
		\end{pmatrix}
	\begin{pmatrix} p_0^\lambda (t) \\  p_1^\lambda (t) \end{pmatrix}.\nonumber
\eea
Note that we have dropped the subscripts $ab,cd$ on $M$ since  
$S_{01} S_{10}=1$. It is convenient to identify the following four rates,
\bea
\begin{pmatrix} \dot p_0^\lambda (t) \\ \dot p_1^\lambda (t) \end{pmatrix} = 
\begin{pmatrix}
	-k_{0\to1}  	& k_{1\to 0}^\lambda 	\\
 	k_{0\to1}^\lambda  & -k_{1\to0}
\end{pmatrix}
\begin{pmatrix} p_0^\lambda (t) \\  p_1^\lambda (t) 
\end{pmatrix}
\label{eq:mat}.
\eea
The two eigenvalues of this matrix are
\bea
\mu^\lambda_{\pm} = - \frac{1}{2} ( k_{0\to1} &+& k_{1\to0}) 
	\pm \frac{1}{2}  \sqrt{ \left( k_{0\to1} + k_{1\to0} \right)^2 
	- 4 \left( k_{0\to1} k_{1\to0}  \nonumber 
	- k_{0\to1}^\lambda k_{1\to0}^\lambda \right) }.
\label{eq:CGFTLS}
\eea
The rate matrix in Eq. (\ref{eq:mat}) is the Liouvillian $\cal L$ and is therefore used to calculate the CGF, 
energy current and noise power. 
%


Recall that the characteristic function ${\cal Z}(\lambda,t)$ is the trace over the counting-field dependent reduced density matrix,
and that it can be used to obtain the CGF,
\bea
{G}(\lambda) = 
\lim_{t \to \infty} \frac{1}{t} \ln \langle I|p^{\lambda}(t)\rangle,
\eea
where $\langle 1|=(1,1)^T$ is the identity vector. In the long-time limit, only
the smallest-magnitude eigenvalue of the Liouvillian survives, and  the CGF is given by
\bea
{G}(\lambda) =  \mu^{\lambda}_+.
\label{eq:CGFTLSs}
\eea
Note that $ \mu^{\lambda=0}_+=0$, which is consistent with the normalization condition of the probabilities. 
We show simulations of $G$ for both ADD and NADD models in Figures \ref{weakC_res}-\ref{TUR} and \ref{niba_res}-\ref{TURniba}, respectively.

Let us now consider the expansion 
\bea
 G(\lambda)= a_1(i\lambda) + a_2(i\lambda)^2+ a_3(i\lambda)^3+ ...
\label{eq:Gexpansion}
\eea
The imaginary part of $G(\lambda)$ is an odd function in $\lambda$. The 
slope of  Im $ G(\lambda)$ around $\lambda=0$ corresponds to the averaged energy current.
Similarly, the real part of $G(\lambda)$ is an even function in $\lambda$. 
The coefficients correspond to the cumulants of energy exchange: energy current ($a_1$), noise power ($2a_2$), skewness, etc. 
\subsection{Energy current and noise power}
\label{subsec-TLS-current-noise}
The energy currents and the corresponding zero-frequency noise powers can be readily obtained by 
taking derivatives of the CGF with respect to the counting-field. 
The first cumulant (current) and second cumulant (noise power) at the $\nu$ bath are given, respectively, by
\bea
\langle J_\nu \rangle &\equiv& \lim_{t \to \infty} \frac{\langle Q_\nu \rangle}{t}
= \frac{\partial {G}(\lambda)}{\partial (i \lambda)}\Big{|}_{\lambda=0}, \nonumber \\
\langle S_\nu \rangle &\equiv& \lim_{t \to \infty} \frac{\langle Q_\nu^2 \rangle-\langle Q_\nu \rangle^2}{t}
= \frac{\partial^2 {G}(\lambda)}{\partial (i \lambda)^2}\Big{|}_{\lambda=0},
\label{eq:current:noise}
\eea
In the long-time limit, we solve Eq. (\ref{eq:mat}) for $\lambda=0$ and find the steady state population
$p_0^{ss}= \frac{k_{1\to0}}{k_{1\to 0}+k_{0\to1}}$, with $p_0+p_1=1$.
We then reach the following  expressions, 
\bea
\langle J_\nu \rangle &=& 
p_0^{ss}\frac{  \partial k_{0\to1}^{\lambda}}  {\partial (i\lambda)}\Big|_{\lambda=0} 
+p_1^{ss}\frac{\partial k_{1\to0}^{\lambda}}{\partial (i\lambda)}\Big|_{\lambda=0}
\nonumber \\
\langle S_\nu \rangle &=& 
 p_0 ^{ss}\frac{\partial^2 k_{0\to1}^{\lambda}}{\partial (i\lambda)^2}\Big|_{\lambda=0} 
+ p_1^{ss} \frac{\partial^2 k_{1\to0}^{\lambda}}{\partial (i\lambda)^2}\Big|_{\lambda=0} 
+\frac{2}{k_{1\to 0} + k_{0\to 1}}
\left[ \left( \frac{\partial k_{0\to1}^{\lambda}}  {\partial (i\lambda)} \frac{\partial k_{1\to0}^{\lambda}}{\partial (i\lambda)} \right)\Big|_{\lambda=0}  - \langle J_\nu \rangle^2 \right].
\label{eq:JG:SG}
\eea
It is significant to  note that these results are general, 
valid for arbitrary bath coupling operator $\hat V_N$. 
In what follows we discuss the so-called  ``thermodynamic uncertainty relation", then calculate the current and noise in the ADD and NADD models.

\subsubsection{Thermodynamics uncertainty relation}
\label{subsubsec-TUR}
In recent studies, an interesting, thermodynamic uncertainty relation was discovered for Markov processes in steady state. It 
connects the steady state averaged current, its fluctuations, and the entropy production rate in 
the nonequilibrium process $\Sigma_Q$ as \cite{TUR1,TUR2,TUR3},
\bea
\frac{\langle S\rangle}{  \langle J\rangle^2} \frac{\Sigma_Q }{ k_B} \geq 2.
\eea
This relation points to a crucial trade off between precision and dissipation:
A precise process with little noise requires high thermodynamic-entropic cost.
For a two-terminal junction with $T_R>T_L$, the entropy production rate is directly related to the current,
$\Sigma_Q=\langle J\rangle \Delta\beta$ with $\Delta \beta=\left( \frac{1}{T_L}-\frac{1}{T_R} \right)$.
The thermodynamic uncertainty relation then reduces to
\bea
\frac{\langle S_R\rangle}{\langle J_R\rangle}\Delta \beta \geq 2k_B.
\label{eq:TUR2}
\eea
In linear response, $\langle J_R \rangle =\kappa \Delta T$, 
$\Delta \beta=\Delta T/T_a^2$, with $2T_a=T_L+T_R$ twice the averaged temperature.
The inequality then reaches the Green-Kubo relation,
$ \langle S_R\rangle = 2k_B T_a^2 \kappa$, with $\kappa$ as the thermal conductance.

In Secs. \ref{subsubsec-TLS-add} and \ref{subsubsec-TLS-nadd} we show that the relation holds in additive, non-additive, symmetric, asymmetric, classical and quantum regimes. 
This follows from the fact that we use assume Markovianity of the dynamics throughout these different cases. 
Beyond that, the inequality can be violated depending on the behavior of high order transport coefficients, 
as we recently proved in Ref. \cite{TURQ}.
\subsubsection{Additive bath interaction model: $\hat V_N = \hat B_L + \hat B_R$}
\label{subsubsec-TLS-add}

In the case of an additive system-bath coupling, $\hat V_{N}=\hat B_L+\hat B_R$, it can be shown that
the rate constants are given by (we count energy at the $R$ terminal) 
%
\bea
k_{0\to 1}^{\lambda} &=& 
\intinf ds \, e^{-i\omega_0s} \left[\langle \hat B_L(s) \hat B_L(0)\rangle + \langle \hat B_R^{\lambda}(s) \hat B_R^{-\lambda}(0)\rangle \right]
\nonumber\\
&\equiv&  k_{0\to 1}^L + k_{0\to 1}^{R, \lambda}\nonumber
\eea
Explicitly, one can readily prove that the counting field rate constant relates to the $\lambda=0$ expression as
$k_{0\to 1}^{R,\lambda}= e^{i\omega_0\lambda}k_{0\to 1}^R$, 
shown previously in Eq. (\ref{eq:MLambdaReln}).
Similarly, $k_{1\to 0}^{R,\lambda}= e^{-i\omega_0\lambda}k_{1\to 0}^R$.
Using these two relations in (\ref{eq:JG:SG}),  the energy current and noise power reduce to
\bea
\langle J_R\rangle &=&\omega_0 \, [p_0^{ss} \, k_{0\to 1}^R - p_1^{ss} \, k_{1\to 0}^R], 
\nonumber \\
\langle S_R\rangle &=& \omega_0^2 \left( p_0^{ss} \, k_{0\to 1}^R + p_1^{ss} \, k_{1\to 0}^R \right)-\frac{2}{k_{1\to 0} + k_{0\to 1}} \left( \langle J_R \rangle^2 + \omega_0^2 k_{0\to 1}^R k_{1\to 0}^R \right).
\label{eq:weak}
\eea
The expression for the current agrees with
previous studies  in which the energy current was defined ``by hand"
in the weak coupling limit [see the discussion below Eq. (\ref{eq:Hsdot})
 \cite{segal-nitzan,segal-master} by constructing an energy current operator \cite{claire}, 
or by performing a classical counting statistics analysis (by resolving the Markovian master equation) \cite{Ren-PRL,segal-nicolin}. 
The noise expression agrees with the zero-frequency limit of Ref. \cite{junjie}.

\begin{figure}
\includegraphics[width=\textwidth]{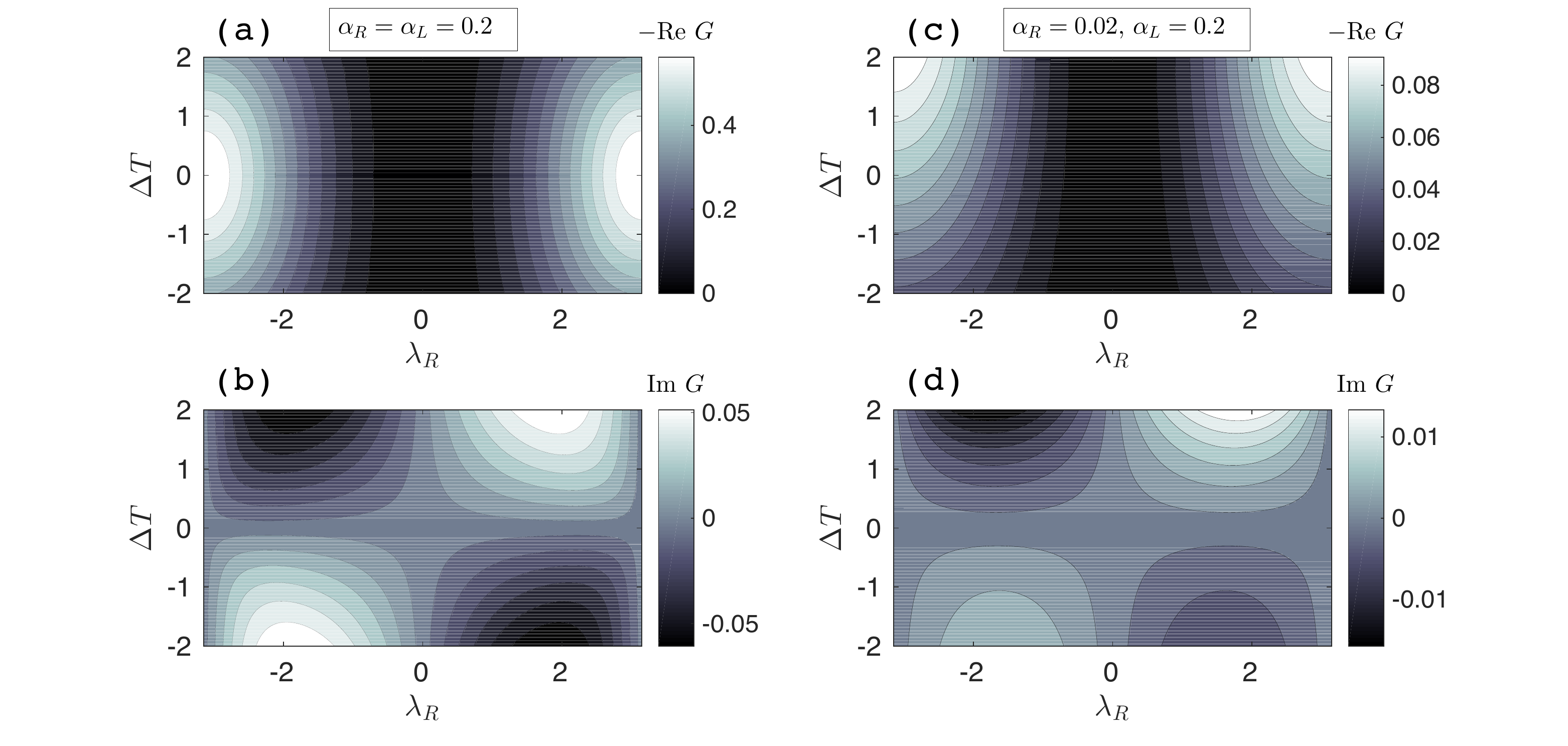} 
\caption{
Cumulant generating function for an unbiased spin system coupled to two harmonic baths with an additive
bath interaction Hamiltonian.
We present Re $G$ and Im $G$ as a function of the counting parameter $\lambda=\lambda_R$ and temperature bias $\Delta T=T_R-T_L$.
(a)-(b) Symmetric junction, $\alpha_L = \alpha_R = 0.2$.
(c)-(d) Asymmetric junction, $\alpha_L = 0.2$, $\alpha_R = 0.02$.
Other parameters are $\omega_0 = 1$, average temperature $T_a = 2$. }
\label{weakC_res}
\end{figure}

Let us consider the nonequilibrium spin-boson model, The system is given by a spin, and the thermal baths are modeled by a collection of noninteracting bosons,
\bea
\hat H = \frac{\Delta}{2} \hat \sigma_z +  \frac{\omega_0}{2} \hat \sigma_x +
\hat \sigma_z \otimes \hat V_{N} +
\sum_{\nu,j}\omega_{\nu,j} \hat b_{\nu,j}^{\dagger}\hat b_{\nu,j}.
\label{eq:HSB}
\eea
Here, 
$\hat \sigma_x$ and $\hat \sigma_z$ are the Pauli matrices as in Eq. (\ref{H-TLS}), 
$\Delta$ is
the energy gap between the spin levels,  and $\omega_0$ is the
tunnelling energy. 
The two reservoirs, $L$ and $R$,  include uncoupled harmonic oscillators,  with
$\hat b_{\nu,j}^{\dagger}$ ($\hat b_{\nu,j}$) as the bosonic creation
(annihilation) operator of the $j$th mode in the $\nu$ reservoir. 
We solve the transport behavior of the model in the separable  bath interaction $\hat V_{N}$ limit,  
$\hat V_{N} = \sum_{\nu,j}\gamma_{\nu,j}(\hat b_{\nu,j}^\dagger + \hat b_{\nu,j})$,
$\gamma_{\nu,j}$ is the system-bath interaction energy. 
For simplicity, let us take $\Delta=0$ in Eq. (\ref{eq:HSB}). 
We then perform a rotation, 
%
$\hat W^{\dagger}\hat \sigma_z \hat W=\hat \sigma_x$, 
$\hat W^{\dagger}\hat \sigma_x \hat W=\hat \sigma_z,$
%
with $\hat W=\frac{1}{\sqrt{2}}(\hat \sigma_x+\hat \sigma_z)$, and receive
the transformed Hamiltonian $\hat H_{ADD}=\hat W^{\dagger}\hat H\hat W$,
\bea
\hat H_{ADD} =   \frac{\omega_0}{2} \hat \sigma_z +
\hat \sigma_x \sum_{\nu,j}\gamma_{\nu,j}(\hat b_{\nu,j}^\dagger + \hat b_{\nu,j}) +
\sum_{\nu,j}\omega_{\nu,j} \hat b_{\nu,j}^{\dagger}\hat b_{\nu,j}.
\label{eq:HSBW}
\eea
This Hamiltonian offers a convenient starting point for a perturbative treatment.
Using the results of Sec. \ref{subsec-TLS-CGF}, 
one can readily write down the CGF of the model (\ref{eq:CGFTLSs}) with 
$k_{1\to 0}^{\nu}=\Gamma_{\nu}(\omega_0)(1+n_{\nu}(\omega_0))$,
and $k_{0\to 1}^{\nu}=\Gamma_{\nu}(\omega_0)n_{\nu}(\omega_0)$, 
the results for which are shown in Fig. \ref{weakC_res}.
%
Here, $n_{\nu}(\omega) = (e^{\beta_\nu \omega}-1)^{-1}$ is the Bose-Einstein distribution function,
 $\Gamma_{\nu}(\omega) = 2\pi \sum_{j \in \nu} \gamma_{\nu,j}^2 \delta(\omega-\omega_{\nu,j})$
is the  spectral density function.
For details, see \cite{segal-nicolin}.
The energy current and noise power at the $R$ terminal are
\begin{eqnarray}
\langle J_R \rangle =
\frac{\omega_0 \,\Gamma_L(\omega_0)\Gamma_R(\omega_0)
\left[ n_R(\omega_0)-n_L(\omega_0)\right]}
{\Gamma_L(\omega_0)[1+2n_L(\omega_0)] + \Gamma_R(\omega_0)[1+2n_R(\omega_0)]},
\nonumber
\end{eqnarray}
\begin{eqnarray}
\langle S_R \rangle =
\frac{ \omega_0^2 \, \Gamma_L(\omega_0)\Gamma_R(\omega_0)
\left[ n_R(\omega_0)\left(1+n_L(\omega_0)\right) + n_L(\omega_0)\left(1+n_R(\omega_0)\right) \right]-2\langle J_R \rangle^2}
{\Gamma_L(\omega_0)[1+2n_L(\omega_0)] + \Gamma_R(\omega_0)[1+2n_R(\omega_0)]}.
\label{eq:JSweak}
\end{eqnarray}
Other approaches for treating this model in a perturbative manner are formulated in e.g. Refs. \cite{wang4,Wu,bijay-majorana}.  

\begin{figure}
\includegraphics[width=\textwidth]{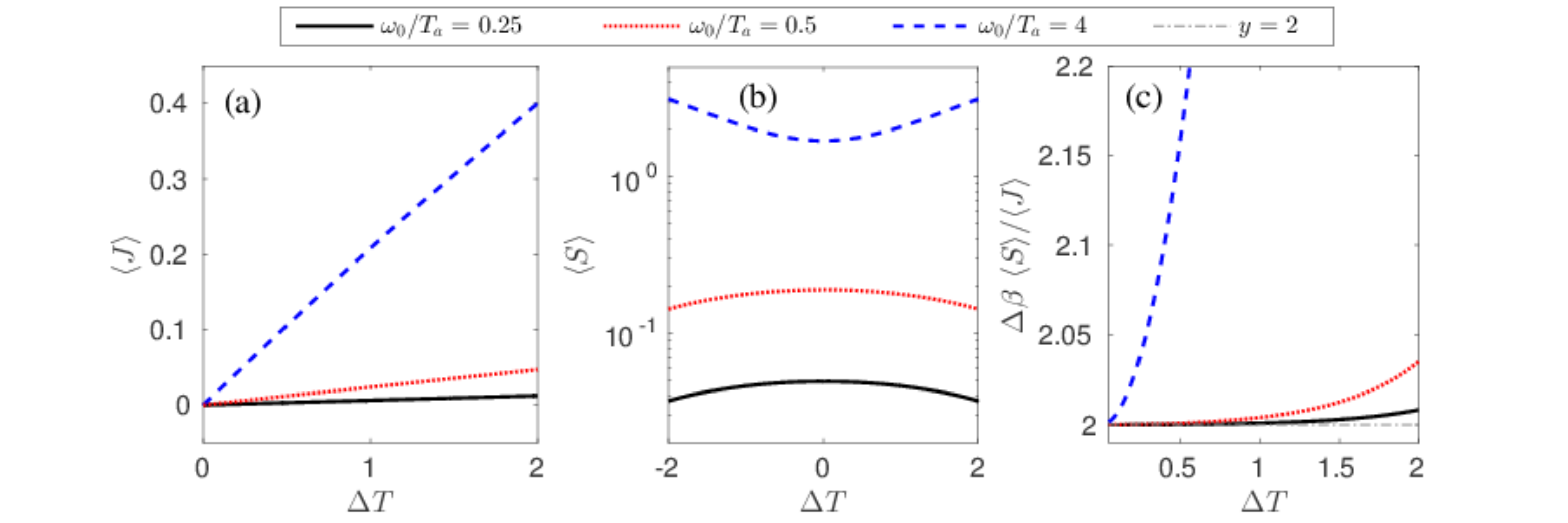} 
\caption{
(a) Current, (b) noise, and (c) the thermodynamic uncertainty relation (\ref{eq:TUR2})
for the additive model (\ref{eq:JSweak}),
$T_a=2$, $\alpha_R=\alpha_L=0.2$,  $\omega_0/T_a=4$ (dashed), $0.5$ (dotted) and $0.25$ (full).
In (c), the bound of $2$ on the ratio $\Delta \beta \, \langle S \rangle/\langle J \rangle$ is shown as a dashed-dotted line.}
\label{TUR}
\end{figure}

We simulate both symmetric and asymmetric junctions in Fig. \ref{weakC_res}, and demonstrate the thermal 
rectification effect, an asymmetry of the current and noise with respect to $\pm \Delta T$. 
In Fig. \ref{TUR} we present the current and its noise for the weak-coupling additive model (\ref{eq:JSweak}). We further
demonstrate that the thermodynamic uncertainty relation, Eq. (\ref{eq:TUR2}), 
is satisfied in both the classical $T_a\gg\omega_0$ and quantum  $T_a\ll\omega_0$ regimes.
It is significant to note that in the high temperature limit the noise 
decreases when increasing $\Delta T$, 
yet the uncertainty relation holds.
Throughout our two-terminal simulations we use an ohmic spectral density  function
$\Gamma_{\nu}(\omega)=\alpha_{\nu}\omega e^{-|\omega|/\omega_\nu}$ with
$\alpha_{\nu}$ as a dimensionless coefficient and $\omega_\nu$ as the cutoff frequency. 
%

\subsubsection{Non-additive bath interaction: $\hat V_N = \hat B_L \otimes \hat B_R$}
\label{subsubsec-TLS-nadd}

We now discuss the NADD model $\hat V_{N}=\hat B_L  \otimes \hat B_R$. 
In this case, the Liouvillian cannot be separated to left and right-reservoir assisted processes.
For a two-level system, the rates, Eq. (\ref{eq:rateMN}), simplify to
\bea
&&
k_{1\to0}^{\lambda}=\intinf ds \, e^{i\omega_0s} \langle \hat B_L(s)\hat B_L(0) \rangle \langle \hat B_R^{\lambda}(s)\hat B_R^{-\lambda}(0) \rangle
=\frac{1}{2\pi}\intinf d\omega \, M_L(\omega_0-\omega) M_R^{\lambda;-\lambda}(\omega).
\nonumber\\
&&
k_{0\to1}^{\lambda}=\intinf ds \, e^{-i\omega_0s} \langle \hat B_L(s)\hat B_L(0) \rangle \langle \hat B_R^{\lambda}(s)\hat B_R^{-\lambda}(0) \rangle
=\frac{1}{2\pi}\intinf d\omega \, M_L(-\omega_0-\omega) M_R^{\lambda;-\lambda}(\omega).
\eea
From Eq. (\ref{eq:MLambdaReln})
we immediately build the energy current and noise power expressions (\ref{eq:JG:SG})
\bea
\langle J_R \rangle 
&=&
\frac{1}{2\pi}\left[
p_0^{ss} \intinf d\omega \, \omega \, M_L(-\omega_0+\omega)M_R(-\omega)
- p_1^{ss} \intinf d\omega \, \omega \, M_L(\omega_0-\omega)M_R(\omega) \right],\nonumber
\eea
\bea
\langle S_R \rangle 
=
\frac{2}{k_{1\to0}+k_{0\to1}}
\left( \frac{1}{4\pi^2} \intinf d\omega \, \omega M_L(\omega_0-\omega) M_R(\omega) 
\intinf d\omega^\prime \, \omega^\prime M_L(-\omega_0-\omega^\prime)M_R(\omega^\prime)
- \langle J_R \rangle^2  \right)
\nonumber \\
+  \frac{p_0^{ss}}{2\pi} \intinf d\omega \, \omega^2 M_L(-\omega_0-\omega) M_R(\omega) + \frac{p_1^{ss}}{2\pi} \intinf d\omega \, \omega^2 M_L(\omega_0-\omega) M_R(\omega)  .
\label{eq:NADDcurrent}
\eea
This energy current can be deduced from Eq. (\ref{eq:J_nadd}). It agrees with 
the expression used ad hoc in Ref. \cite{segal-nitzan} and with
the result of Ref. \cite{segal-nicolin}, where counting was done
by resolving the population dynamics.
It can be rationalized by interpreting $\omega \, M_L(\omega_0-\omega) M_R (\omega)\, p_1^{ss}$ as
a relaxation process with the energy $\omega$ emitted to the
$R$ reservoir, and the amount of $\omega_0-\omega$ disposed into the $L$ reservoir. 
The baths thus work cooperatively. 
A similar interpretation holds for the other term.
It is significant to
note that this expression has been achieved under relatively
general conditions, without specifying the nature of the bath coupling operator, 
aside from assuming it is in a product form.

\begin{figure}
\includegraphics[width=\textwidth]{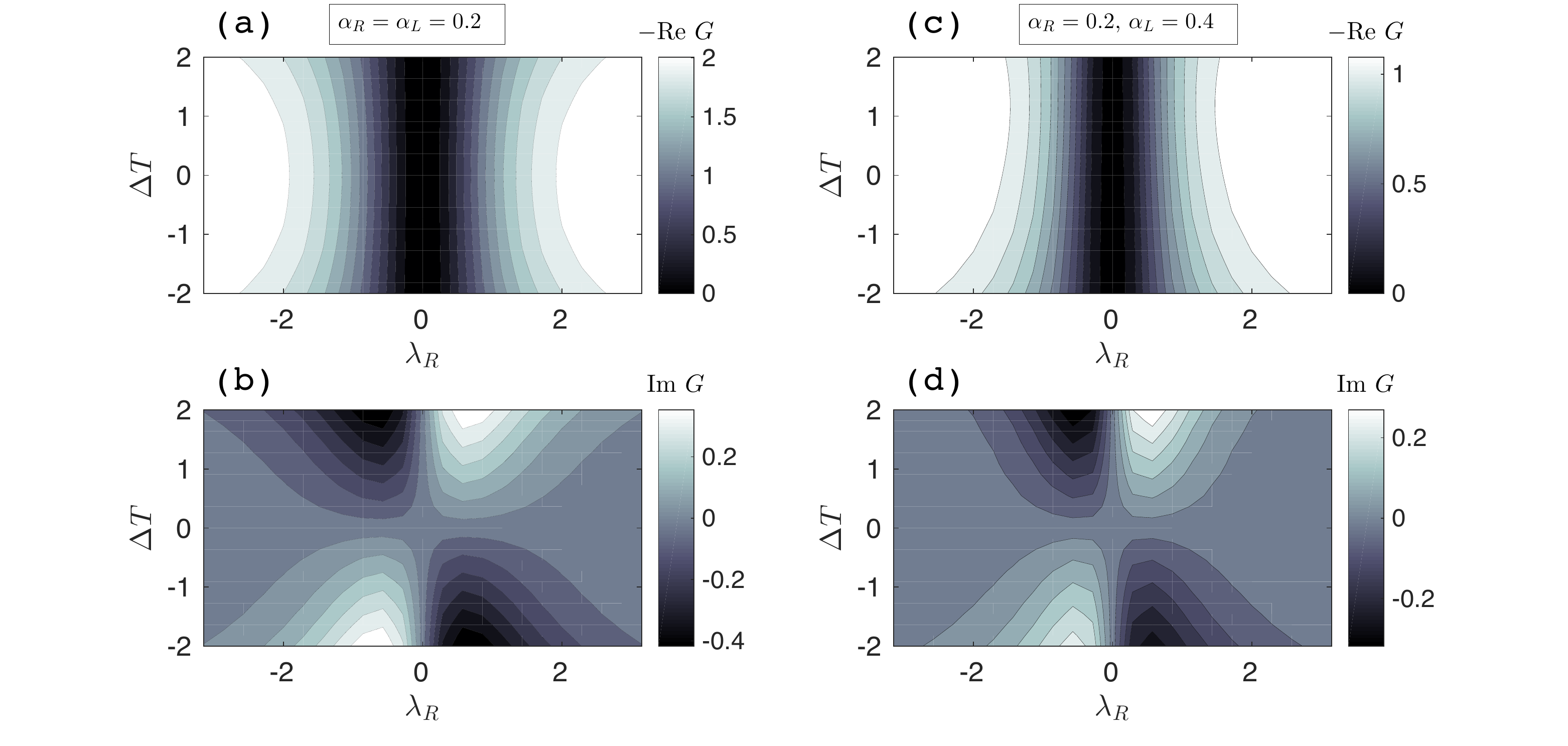} 
\caption{Cumulant generating function for 
the polaronic NADD model (\ref{eq:HSBp}) with
 $\omega_0 = 1$, average temperature $T_a = 2$, $\omega_\nu = 10$, 
(a)-(b) symmetric junction with $\alpha_R=\alpha_L=0.2$. 
(c)-(d)  asymmetric junction $\alpha_R=0.2$, $\alpha_L=0.4$.
}
\label{niba_res}
\end{figure}
\begin{figure}
\includegraphics[width=\textwidth]{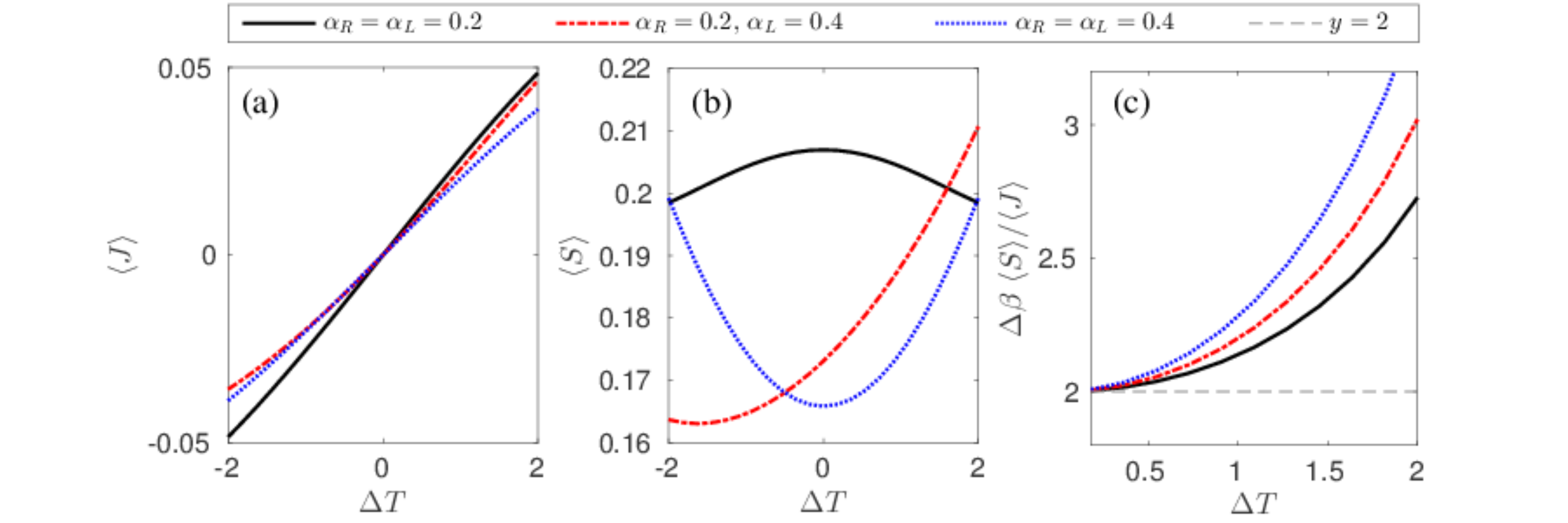}   
\caption{
(a) Current, (b) noise, and (c) the thermodynamic uncertainty relation (\ref{eq:TUR2}) with respect to $\Delta T=T_R-T_L$ 
for the polaronic NADD model (\ref{eq:HSBp}).
Parameters are the same as in Fig. \ref{niba_res}.
In (c), the lower bound of $2$ is shown as a dashed line.} 
\label{TURniba}
\end{figure}

Back to Eq. (\ref{eq:HSB}),
we set $\hat V_{N} = \sum_{\nu,j}\gamma_{\nu,j}(\hat b_{\nu,j}^\dagger + \hat b_{\nu,j})$ 
and interrogate the strong-system bath coupling limit
by performing the polaron transformation  \cite{Mahan},
$\hat H_P=\hat P^{\dagger}\hat H\hat P$ with
$\hat P=e^{i \hat \sigma_z\hat\Omega/2}$,
\bea
\hat H_{NADD} = \frac{\Delta}{2} \hat \sigma_z +
 \frac{\omega_0}{2}\left( \hat \sigma_+ e^{i\hat\Omega} + \hat \sigma_- e^{-i\hat\Omega} \right)
+\sum_{\nu,j}\omega_{\nu,j} \hat b_{\nu,j}^{\dagger}\hat b_{\nu,j},
\label{eq:HSBp}
\eea
where $\hat \sigma_{\pm}=\frac{1}{2}(\hat \sigma_x\pm i \hat \sigma_y)$, 
$\hat\Omega=\sum_{\nu}\hat\Omega_{\nu}$,
and $\hat\Omega_{\nu}=2i\sum_{j}\frac{\gamma_{\nu,j}}{\omega_{\nu,j}}(\hat b_{\nu,j}^{\dagger}-\hat b_{\nu,j})$.
The tunneling splitting $\omega_0$ is thus dressed by a product of shift operators,
 $\Pi_{\nu=L,R} \exp\left[-2\sum_{j}  \frac{\gamma_{\nu,j}}{\omega_{\nu,j}}(\hat b_{\nu,j}^{\dagger}-\hat b_{\nu,j})  \right]$, a
 non-additive  bath coupling model.
The CGF of the model is given by Eq. (\ref{eq:CGFTLS}), with the correlation functions 
\bea
M_{\nu}(t) =\frac{\omega_0}{2}e^{-{\cal{Q}}_{\nu}(t)},
\eea
where the real and imaginary parts, ${\cal{Q}}_{\nu}(t)={\cal{Q}}_{\nu}'(t)+i{\cal{Q}}_{\nu}(t)''$ fulfill
\bea
{\cal{Q}}''_{\nu}(t)& = &  2\int_{0}^{\infty} \frac{\Gamma_{\nu}(\omega)}{\pi\omega^2}\sin(\omega t)\,d\omega,
\nonumber\\
{\cal{Q}}'_{\nu}(t)& = & 2\int_{0}^{\infty}\frac{\Gamma_{\nu}(\omega)}{\pi\omega^2}[1-\cos(\omega t)] [1+2n_{\nu}(\omega)] \,d\omega.
\label{eq:QSB}
\eea
The energy current of the model is given by Eq. (\ref{eq:NADDcurrent}), which can be solved analytically in e.g. the so-called Marcus (strong coupling high temperature) limit \cite{segal-nitzan,segal-nicolin}. 
Extensions to this result were discussed in Refs. \cite{Ren1,Ren2}, 
going beyond the assumption of $\langle \hat V_{N}\rangle=0$.
Other studies had employed the polaron picture as a starting point for higher order 
perturbative treatments \cite{aslangul, Li-majorana}.
We display the CGF in Fig. \ref{niba_res},  and exemplify the current and its noise in Fig. \ref{TURniba},
exposing a thermal diode effect.
Panel (a) in Fig. \ref{TURniba} also reveals the turnover effect of strong system-bath coupling, as
a symmetric junction with $\alpha_\nu = 0.4$ shows a smaller current than with $\alpha_\nu = 0.2$.
The behavior of the current noise is quite interesting as it may increase or decrease with $\Delta T$.
%

%
\section{Quantum absorption refrigerator}
\label{sec-engine}

Based on  the formalism organized in this paper, we are now ready to derive the working equations of the qubit-refrigerator,
which was described in Ref. \cite{Anqi}.
Most interestingly, a non-additive bath coupling model can realize the qubit-QAR,
but this function is missing in the additive case.

An autonomous absorption refrigerator transfers thermal energy from a cold ($C$) bath to a hot ($H$) bath without an input power
by using thermal energy provided from a so-called work ($W$) reservoir, where ${T}_{W}> {T}_{H}> {T}_{C}$.
A common design of a QAR consists of a three-level system, 
where each transition between a pair of levels is coupled to only one of the three thermal baths \cite{QAR,Linden11,joseSR}. 
By tuning the level spacing of the three-level system, one can operate the system as a refrigerator, 
extract energy from the cold bath, and dump it into the hot one.
The three-level QAR operates optimally when system-bath coupling is weak. 

In a recent study \cite{Anqi}, we demonstrated that the smallest system, a qubit, is incapable of operating as 
a refrigerator when the baths are coupled additively to the system. 
The energy current from the cold bath in this case can be derived from the expressions in Sec. \ref{subsubsec-TLS-add}, resulting in 
\bea
	\langle J_C\rangle = -\frac{\omega_0}{k_{0\to1}+k_{1\to0}} \left[ k^{C}_{1\to0} k^{H}_{1\to0} \left( \frac{k^{H}_{0\to1}}{k^{H}_{1\to0}} - \frac{k^{C}_{0\to1}}{k^{C}_{1\to0}} \right) 
	+ k^{C}_{1\to0} k^{W}_{1\to0} \left( \frac{k^{W}_{0\to1}}{k^{W}_{1\to0}} - \frac{k^{C}_{0\to1}}{k^{C}_{1\to0}} \right) \right].
\label{eq:JC_weak}
\eea
Using the detailed balance relation and the fact that $(e^{-\beta_{H,W} \omega_0}-e^{-\beta_C \omega_0})>0$, 
we conclude that $\langle J_C\rangle<0$ regardless of the details of the model. 
Equation (\ref{eq:JC_weak}) shows that under the additive model at weak coupling, 
every two reservoirs exchange energy independently and thus thermal energy always flows to the colder bath, 
and the chiller performance is unattainable. 
In contrast, in Ref. \cite{Anqi} we showed that by coupling the system in a non-additive manner to three heat baths 
which are spectrally structured, a QAR could be achieved. 
Moreover, we demonstrated that the system could reach the Carnot bound
if the reservoirs were characterized by a single frequency component \cite{Anqi}.

The groundwork of the qubit-QAR of Ref. \cite{Anqi} is presented in this paper.
The design is based on a three-bath non-additive interaction form, 
$\hat V_N=\hat B_H \otimes \hat B_C \otimes \hat B_W$ with $\hat S=\hat \sigma_x$.
The three baths are characterized by different spectral properties and maintained at different temperatures. 
Following Sec. \ref{sec-TLS},  
we perform an FCS analysis for each bath, to obtain the current directed to the system from each terminal. 
For simplicity, we count energy only from $C$ using the counting parameter $\lambda$.
Transitions between the system's states are dictated by the rate constants 
\bea
k_{1\to0}^\lambda &=& \frac{1}{(2\pi)^2}\intinf d\omega \, d\omega' M_C(\omega) M_H(\omega') M_W(\omega_0-\omega-\omega'), 
\nonumber \\
k_{0\to1}^\lambda &=&\frac{1}{(2\pi)^2}\intinf d\omega \, d\omega' M_C(\omega) M_H(\omega') M_W(-\omega_0-\omega-\omega').
\eea
The energy current, from the cold bath to the system, is given by Eq. (\ref{eq:JG:SG}), 
\bea
\langle J_C \rangle =
\frac{1}{(2\pi)^2}\Big[ \, p_0^{ss} \intinf d\omega \, d\omega' \, 
\omega M_C(-\omega)M_H(-\omega') M_W(-\omega_0+\omega+\omega') \nonumber \\
-p_1^{ss} \intinf d\omega \, d\omega'  \, \omega M_C(\omega)M_H(\omega')M_W(\omega_0-\omega-\omega') \Big].
\eea
Analogous expressions can be written for $\langle J_H \rangle$ and $\langle J_W \rangle$. 
We also calculate the noise power using Eq.  (\ref{eq:JG:SG}) and arrive at,
\bea
\langle S_C \rangle =
\frac{2}{k_{1\to0}+k_{0\to1}}
\Big( \frac{1}{(2\pi)^4} \intinf d\omega \, d\omega' \omega M_C(\omega) M_H(\omega^\prime) M_W(-\omega_0-\omega-\omega^\prime) \nonumber \\
\times \intinf d\omega \, d\omega' \omega M_C(\omega) M_H(\omega') M_W(\omega_0-\omega-\omega')
- \langle J_C \rangle^2  \Big)\nonumber \\
+  \frac{p_0^{ss}}{(2\pi)^2} \intinf d\omega \, d\omega' \omega^2 M_C(\omega) M_H(\omega^\prime) M_W(-\omega_0-\omega-\omega^\prime) \nonumber \\
+ \frac{p_1^{ss}}{(2\pi)^2} \intinf d\omega \, d\omega' \omega^2 M_C(\omega) M_H(\omega^\prime) M_W(\omega_0-\omega-\omega^\prime) .
\label{eq:SC}
\eea

Based on these expressions, we study the
operation of an absorption refrigerator.
As was demonstrated in Ref. \cite{Anqi}, the design is quite robust, and
a qubit-chiller can be realized with the bath correlation function
$M_\nu(\omega)$ taking a variety of forms, such as a Heaviside box function.
Another convenient form is a bimodal Gaussian function \cite{Anqi},
\bea
M_\nu(\omega)
= \frac{1}{\sigma_\nu}
\left( e^{ -(\omega-\theta_\nu)^2/2 \sigma_\nu^2 }
+ e^{ -(\omega+\theta_\nu)^2/2 \sigma_\nu^2 }
e^{ \beta_\nu \omega } \right),
\label{eq:bimodal}
\eea
which by construction satisfies the detailed balance relation.
Here, $\theta_{\nu}$ is a central frequency that characterizes the spectrum, $\sigma_{\nu}$ is a width parameter.
When $\sigma\to0$, the function collapses to a Dirac delta function---at positive and negative frequencies.
By further setting $\theta_C+\theta_W=\theta_H$, one can analytically prove that the QAR can approach the Carnot bound \cite{Anqi}.
In this special limit, the cooling window is defined by
\bea
\left(\frac{T_W-T_H}{T_W-T_C}\right)\frac{T_C}{T_H}
\geq \frac{\theta_C}{\theta_H},
\label{eq:cool-op}
\eea
which precisely corresponds to the cooling condition as obtained for a three-level or three-qubit QAR
---analyzed with a Markovian
master equation with additive dissipators~\cite{Linden11,QAR,joseSR}.

\begin{figure}
\includegraphics[width=15cm]{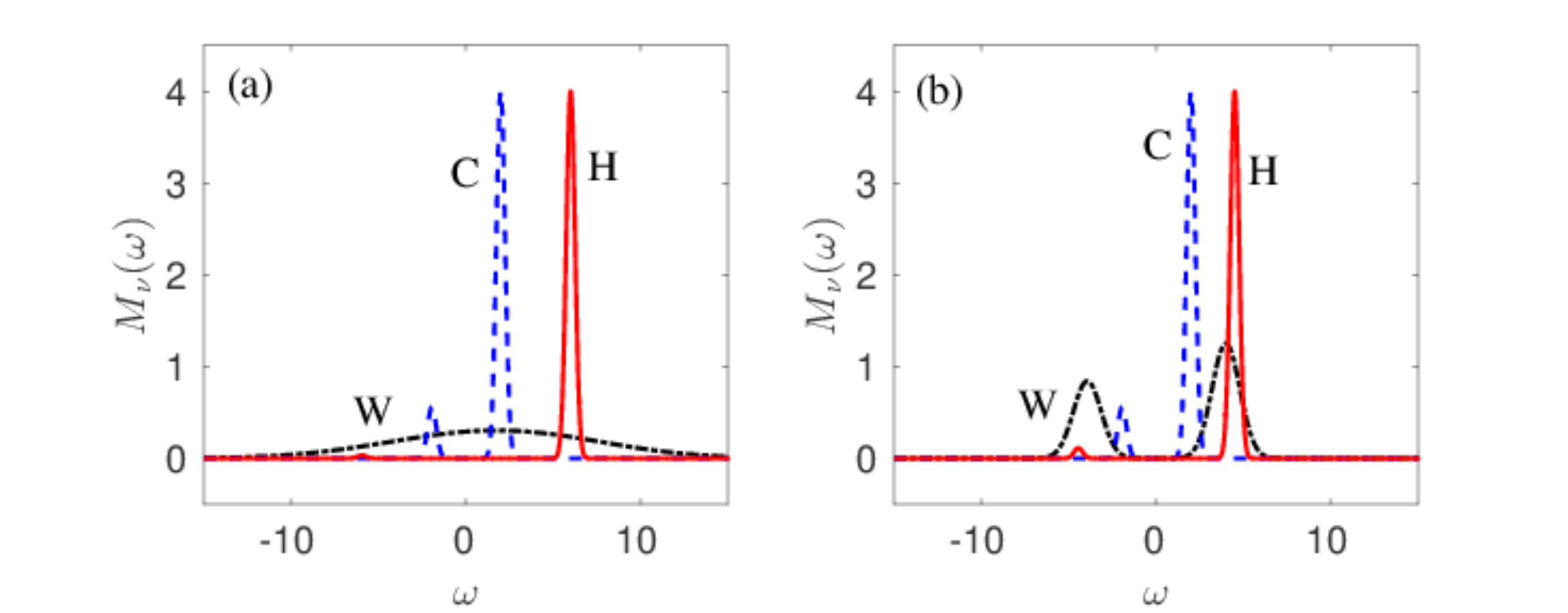} 
\caption{Bimodal bath correlation functions: 
$M_W(\omega)$ (dashed-dotted), $M_H(\omega)$ (full) and $M_C(\omega)$ (dashed) with
$\sigma_C=0.25$, $\sigma_H=0.25$. 
(a) Resonant-broadband $W$ bath, $\theta_C=2$, $\theta_W=4$, $\theta_H=6$, $\sigma_W=5$.
(b) Off-resonant model, $\theta_C=2$, $\theta_W=4$, $\theta_H=4.5$, $\sigma_W=0.8$. }
\label{QAR1}
\end{figure}

\begin{figure}[htbp]
\centering
\includegraphics[width=\textwidth ]{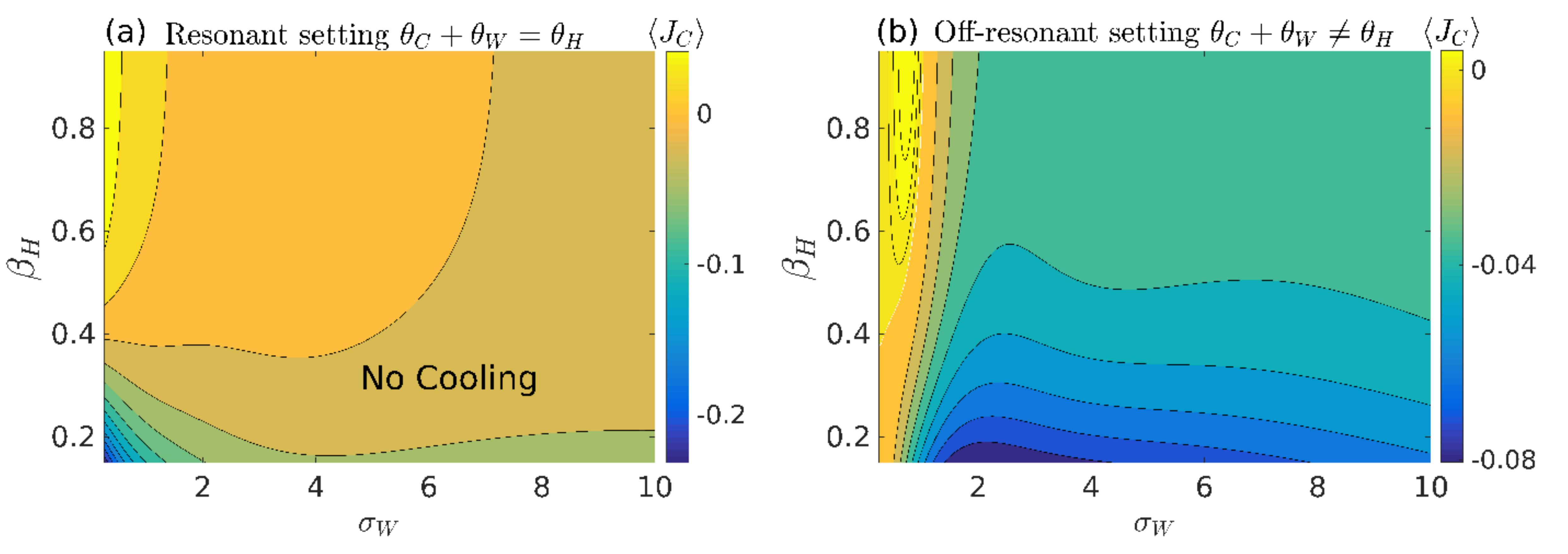} 
\caption{Cooling current in the QAR
as a function of the inverse temperature $\beta_H$ and width parameter
$\sigma_W$. 
(a) Resonance setting, $\theta_H=6$, $\theta_C=2$, $\theta_W=4$. 
(b) Off-resonance setting, $\theta_H=4.5$, $\theta_C=2$, $\theta_W=4$. 
Other parameters are $\beta_W = 0.1$, $\beta_C=1$, $\sigma_H=\sigma_C=0.25$.
}
\label{QAR2}
\end{figure}

We now demonstrate a cooling performance over a broad range of parameters, beyond the resonance condition
and the Delta function (highly engineered-bath) limit analyzed in Ref. \cite{Anqi}.
In Fig. \ref{QAR1}, we present the bimodal functions $M_{\nu}(\omega)$ for the three baths, and
consider two cases: (a) We maintain the resonance condition $\theta_C+\theta_W=\theta_H$, 
but depart from the standard setting by using an un-structured work reservoir, employing a very broad function $M_W(\omega)$. 
(b) We structure the reservoirs with a smaller width $\sigma_W$, but do not
satisfy the resonance condition.
The current $\langle J_C \rangle$ is presented in Fig. \ref{QAR2} for these two cases.
Both setups  achieve cooling $\langle J_C \rangle>0$,  demonstrating that the refrigerator is robust for a variety of setups.
It survives even when the work reservoir is practically structureless. 
As well, it can operate beyond the strict resonance condition by tuning the width $\sigma_W$.

It is useful to note that according to Eq. (\ref{eq:cool-op}), we identify the optimal cooling windows for panels (a) and (b) as 
$\beta_H\geq 0.4$ and $\beta_H\geq 0.5$, respectively. 
Indeed these inequalities serve as good estimates for the cooling window when $\sigma_W$ is small.
Finally, we recall that as long as one of the bath correlation functions is structureless in frequency domain
(decays fast in time domain), the overall correlation function, which is a product of the individual time correlation functions 
(\ref{eq:Mprod}) dies quickly, justifying the Markov assumption that is underlying this analysis. 
%

\section{Summary}
\label{sec-summary}

We provided the theoretical groundwork for the bath-cooperative qubit-QAR of Ref. \cite{Anqi}.
More generally, we presented here a rigorous, thermodynamically consistent treatment of quantum energy transport in small systems 
while going beyond the standard, additive interaction form. 
Using a full-counting statistics approach, we derived a counting-field dependent Redfield-type equation.
After the secular approximation, 
we obtained the cumulant generating function of the model,
particularly, the averaged energy current and its noise power for both additive and non-additive models, 
on the same footing.
Our work illustrated that by studying non-additive models within a weak coupling method one could capture strong-coupling (e.g. multi-phonon) effects
that are not realized with an additive weak-coupling treatment.

We exemplified our results on a qubit system: We studied the transport behavior of a two-terminal setup, 
and the cooling function of a three-terminal quantum absorption refrigerator model. 
In the latter case, we demonstrated that the cooling function was robust, surviving for a broad range of 
baths' parameters.

Applications explored in this work rely on the secular approximation 
(decoupling population and coherence dynamics). 
Nevertheless, the formalism could be used beyond that. 
The counting-field dependent Redfield equation could be employed to examine the 
role of quantum coherences in energy transport behavior within multi-level quantum systems, and in the operation of 
quantum heat engines. Analysis of composite heat engine models, e.g., made of several qubits \cite{jose15,Johal},
is left for future studies. It is furthermore essential to examine the behavior of the qubit 
absorption refrigerator with numerically exact approaches
and confirm our predictions.  
 
A full-counting statistics analysis of heat exchange is paramount for various reasons. 
Fundamentally, testing whether the cumulant generating function satisfies the fluctuation relation
immediately reports on the thermodynamic consistency of the employed method.
Practically, one can calculate the current and its noise far from equilibrium---directly from the  cumulant generating function.
While most studies in quantum transport and quantum thermodynamics are focused on 
the calculation of the averaged energy current within a device,
evaluating the current noise is critical for estimating the precision of a process \cite{bayan}. 
This nonequilibrium fluctuation-dissipation trade off is captured by the thermodynamic uncertainty relation, 
which is satisfied within our Markovian QME treatment.
In future work we will interrogate this relation in strongly-coupled quantum heat machines, beyond the Markovian limit.

\section*{Acknowledgments}
DS acknowledges support from an NSERC Discovery Grant and the Canada Research Chair program.
The work of HMF was supported by NSERC CGS-M and NSERC PGS-D programs.
BKA acknowledges support from the CQIQC at the University of Toronto.


\renewcommand{\theequation}{A\arabic{equation}}
\setcounter{equation}{0}  

\section*{Appendix A: Derivation of the counting-field dependent Redfield Equation}
\label{app:AppA}

The formally exact Nakajima-Zwanzig 
generalized quantum master equation (GQME) \cite{Nitzan} can be  extended to include
counting information, as we now explain.
We consider an open quantum system Hamiltonian,  $\hat H = \hat H_S + \hat H_B+ \hat H_{int}$, with the total density operator
$\rho(t)$.
In the absence of counting parameters, the quantum Liouville equation can be written as (Schr\"{o}dinger picture),
\bea
\dot \rho(t)&=& -i[\hat H,\rho(t)] 
\nonumber\\
&=&  -i{\cal L} \rho,
\eea
where ${\cal L}_{mnm'n'}= H_{mm'}\delta_{nn'}-\delta_{mm'} H_{n'n}$.
Following Eq. (\ref{eq:rho_com}), we  generalize this standard definition to our needs, 
\bea
\dot \rho^{\lambda}(t)&=& -i\hat H^{-\lambda}\rho^{\lambda}(t) + i\rho^{\lambda}(t) \hat H^{\lambda} 
\nonumber\\
&\equiv&  -i{\cal L}^{\lambda} \rho^{\lambda}(t),
\eea
where
\bea
{\cal L}_{mnm'n'}^{\lambda}= H^{-\lambda}_{mm'}\delta_{nn'}-\delta_{mm'} H^{\lambda}_{n'n}.
\label{eq:Llam}
\eea
The counting-field dependent Hamiltonian $\hat H^{\pm \lambda}$ is defined below Eq. (\ref{eq:U_xi}).
Using projection operators, we proceed and write down the exact  GQME \cite{Breuer,Nitzan},
\bea
\dot \sigma^{\lambda}(t)  = -i{\cal L}_S \sigma^{\lambda}(t) - \int_0^t ds {\cal K}^{\lambda}(s) \sigma^{\lambda}(t-s),
\label{eq:NZ}
\eea
with the kernel 
\bea
{\cal K}^{\lambda}(t) = {\rm Tr}_{\textrm B} 
\left[ {\cal L}_{int}^{\lambda} e^{-i \hat{\cal Q}{\cal L}^{\lambda}t } \hat {\cal Q} {\cal L}^{\lambda}_{int} \rho_{\textrm B}(0) \right].
\label{eq:kernel}
\eea
Here the system Liouvillian is  
${\cal L}_S (\cdot)=[\hat H_{S}, \cdot]$ and 
${\cal L}^{\lambda}_{int}$  is defined as in Eq. (\ref{eq:Llam}).
To arrive at this GQME
we  assumed a factorized initial condition  $\rho(0)=\sigma(0)\otimes\rho_{\textrm{B}}(0)$,
and that $\langle \hat H^{\pm \lambda}_{int} \rangle \equiv {\rm Tr}_{\textrm{B}}  \left[\hat H^{\pm \lambda}_{int}(0) \rho(0)\right]=0$.
The projection operators $\hat {\cal Q}=1-\hat {\cal P}$ are chosen such that we project out the degrees of freedom
of the baths, $\hat {\cal P}(\cdot)= \rho_{\textrm B}(0) {\rm Tr}_{\textrm B}(\cdot)$.
Eqs. (\ref{eq:NZ})-(\ref{eq:kernel}) are simple generalizations
of the exact Nakajima-Zwanzig quantum master equation to include counting information. 
Approximate expressions, developed to solve this formal result, see e.g. Ref. \cite{Geva}, 
can be readily generalized to the present case.
In particular,  in the weak coupling limit the exact kernel reduces to
\bea
{\cal K^\lambda}(t) \sim {\rm Tr}_{\textrm B}\left[ {\cal L}_{int}^{\lambda} e^{-i{\cal L}_S t}e^{-i{\cal L}_B t} {\cal L}_{int}^{\lambda} \rho_{\textrm B}(0) \right].
\label{eq:Kweak}
\eea
Once further performing the Markovian approximation,
Eq. (\ref{eq:NZ}) with (\ref{eq:Kweak}) precisely recovers the 
Redfield equation with counting parameters (\ref{eq:Redfield_counting}).

In what follows, we develop the counting-field dependent Redfield equation in more details 
by following the standard derivation of the second order Markovian master equation \cite{Breuer}.
In doing so, we highlight on the different approximations involved and elaborate on their impact
on the range of systems that can be simulated with this approach.

We begin with Eq. (\ref{eq:rho_com}) and follow the standard derivation of the second order Markovian master equation \cite{Breuer}:
We integrate this equation and insert the formal solution to $\rho^{\lambda}(t)$ back into Eq. (\ref{eq:rho_com}).
We trace out the bath degrees of freedom,  and obtain the reduced density matrix
$\sigma^\lambda(t) \equiv  {\rm Tr}_{\textrm{B}} \left[ \rho^\lambda (t)\right]$, satisfying
\bea
\dot \sigma^\lambda (t)
= &-& \int_0^t  {\rm Tr}_{\textrm{B}}  \left[ \hat H^{-\lambda}_{int}(t)\hat H^{-\lambda}_{int}(t-s)\rho^\lambda (t-s)\right] \, ds
                        + \int_0^t  {\rm Tr}_{\textrm{B}}  \left[ \hat H^{-\lambda}_{int}(t) \rho^\lambda (t-s) \hat H^{\lambda}_{int}(t-s)\right] \, ds  \nonumber\\
                        &+& \int_0^t  {\rm Tr}_{\textrm{B}}  \left[ \hat H^{-\lambda}_{int}(t-s) \rho^\lambda (t-s) \hat H^{\lambda}_{int}(t)\right] \, ds
                        - \int_0^t  {\rm Tr}_{\textrm{B}}  \left[ \rho^\lambda (t-s) \hat H^{\lambda}_{int}(t-s)\hat H^{\lambda}_{int}(t)\right] \, ds.
\label{eq:Asigma1}
\eea
Note the critical assumption made here,
namely, $\langle \hat H^{\pm \lambda}_{int} \rangle \equiv {\rm Tr}_{\textrm{B}}  \left[\hat H^{\pm \lambda}_{int}(0) \rho(0)\right]=0$.
This restricts the structure of the interaction operator, or the range of temperatures and interaction energies for which results
are valid.
For example, in the nonequilibrium spin-boson model under the polaron transformation,
see Refs. \cite{segal-nicolin,FCS-strong,nazim},  $\langle \hat H^{\pm \lambda}_{int} \rangle$
decays exponentially with temperature and system-bath coupling, thus the method as described here
is valid in a high temperature and strong-coupling regime.
More generally, one could re-organize the total Hamiltonian as
$\hat H  = \hat H_S + \langle \hat H_{int} \rangle  + \sum_{\nu=1}^N \hat H_\nu + \hat H_{int}-  \langle \hat H_{int} \rangle$,
and derive a quantum master equation  in the eigenbasis of the
new system Hamiltonian, $\left[\hat H_S + \langle \hat H_{int} \rangle\right]$, see e.g. Refs. \cite{Ren1,Ren2}. 
The resulting QME correctly describes both the high and low temperature limits of the spin-boson model (within that order in perturbation theory).
In this work,
however,
with the objective of keeping our presentation and results simple and transparent, we ignore the
averaged interaction term.

We assume that the initial density matrix takes a factorized form, $\rho(0)=\sigma(0)\otimes\rho_{\textrm{B}}(0)$,
$\rho_{\textrm{B}}(0) =  \prod_{\nu=1}^N \rho_\nu$
with the reservoirs prepared in a canonical state at inverse temperature $\beta_{\nu}$,
$\rho_{\nu}(0)=\exp[-\beta_{\nu} \hat H_{\nu}]/Z_{\nu}$, and $Z_{\nu}$ as the partition function for the $\nu$ bath.
Memory effects from an initially correlated unfactorized initial density matrix are relevant on short timescales
in the weak coupling approximation, and do not affect long-time behavior once these correlations have died out \cite{oppenheim}. 
We proceed with the Born approximation, $\rho^{\lambda}(t-s)\sim \sigma^{\lambda}(t-s)\otimes\rho_{\textrm{B}}(t-s)$,  
which is justified because the system has little effect on the much larger baths.
We assume the reservoirs do not change from their initial state over time,  $\rho_\nu\approx \rho_\nu(0)$.

To progress from Eq. (\ref{eq:Asigma1}), we recall that $\hat H^{\lambda}_{int} = \gamma \sum_p \hat S^p \otimes \hat V_{N}^{p, \lambda}$
and define the two-time bath correlation functions as
$\langle \hat V_{N}^{p,\lambda}(t-s) \hat V_{N}^{p',\lambda'}(t) \rangle \equiv  {\rm Tr}_{\textrm{B}} [ \hat V_{N}^{p,\lambda}(t-s) \hat V_{N}^{p',\lambda'}(t)  \rho_{\textrm{B}}  ]$.
Note that when operators appear with the same sign for the counting-field,
it cancels out,
\bea
\langle \hat V_{N}^{p,\pm \lambda}(t)  \hat V_{N}^{p',\pm\lambda}(t-s)  \rangle
 &=& {\rm Tr}_{\textrm{B}} \left[ e^{\pm i \sum_\nu \lambda_\nu \hat H_\nu/2} \, \hat V_N^p (t) \, e^{\mp i\sum_\nu \lambda_\nu \hat H_\nu/2}  e^{\pm i\sum_\nu \lambda_\nu \hat H_\nu/2} \, \hat V^{p'}_N (t-s) \, e^{\mp i\sum_\nu \lambda_\nu \hat H_\nu/2} \rho_{\textrm{B}} \right]
 \nonumber \\
&=&\langle \hat V_{N}^p(t)  \hat V_{N}^{p'}(t-s)  \rangle = \langle \hat V^p_{N}(s)  \hat V^{p'}_{N}(0)  \rangle.
\eea
The second equality arises due to the cyclic properties of the trace operation and the fact that $\hat H_\nu$ and $\rho_{\textrm{B}}$ commute.
The last equality relies on the fact that the bath is stationary.
We define this coupling correlation function as,
\bea
M^{p,\lambda;p^\prime, \lambda^\prime}_{N; \, {ab,cd}}(s)
\equiv
\gamma^2 \, S^p_{{ab}} S^{p^\prime}_{{cd}} \, \langle \hat V_{N}^{p,\lambda}(s) \hat V_{N}^{p^\prime,\lambda^\prime}(0) \rangle,
\label{eq:AMs}
\eea
and its half Fourier transform,
\bea
R_{N; \,{ab,cd}}^{p,\lambda;p^\prime,\lambda^{\prime\,(+)}}(\omega)
\equiv \int_0^\infty e^{i\omega s} M^{p,\lambda;p^\prime,\lambda^\prime}_{N;\,{ab,cd}}(s) \, ds , \,\,\,\,\,\,\,\,\,\,\,\,\,
R_{N;\,{ab,cd}}^{p,\lambda;p^\prime,\lambda^{\prime\,(-)}}(\omega)
\equiv \int_{-\infty}^0 e^{i\omega s} M^{p,\lambda;p^\prime,\lambda^\prime}_{N;\,{ab,cd}}(s) \, ds.
\label{eq:ARw}
\eea
The terms $S^p_{{ab}}$ and $S^{p^\prime}_{{cd}}$ are the matrix elements of the $\hat S^p$ and $\hat S^{p^\prime}$ operators respectively, in the eigenbasis of the system Hamiltonian $\hat H_S$.

We now study Eq. (\ref{eq:Asigma1}) in the Markovian limit,
assuming the dynamics of the bath is much faster than those of the system.
Firstly, we replace
$\sigma^\lambda(t-s)\rightarrow\sigma^\lambda(t)$ since in the weak coupling limit,
the interaction parameter $\gamma$ is small.
Recall that Eq. (\ref{eq:Asigma1}) is written in the interaction representation.
Second, we assume that bath correlation functions decay to small values
much faster than any characteristic system timescale.
This allows us to take the upper limit of the integrals to infinity,
where beyond $t$, the contributions from the integrand due to the bath-bath correlation function,
having decayed quickly, are negligible.
The validity of the Markovian master equation has been extensively studied \cite{Breuer,plenio2} 
and is valid for our purposes as long as $\gamma$ is small and the (broadband) bath is characterized by resonant modes with the system.

These approximations result in the Markovian QME, the counting-field dependent Redfield-type equation \cite{Nitzan},
\bea
\dot \sigma_{nm}^\lambda (t) = -i[\hat H_S,\sigma(t)]_{nm} +
         \sum_{p,q,j,k} \Big[
&-& \sigma_{km}^\lambda(t) R^{p,q^{(+)}}_{N;\,{nj,jk}}(E_{k,j})
 -  \sigma_{nj}^\lambda(t) R^{p,q^{(-)}}_{N;\,{jk,km}}(E_{j,k})
\nonumber \\
&+& \sigma^\lambda_{jk}(t)  \left( R^{q,\lambda,p,-\lambda^{(-)}}_{N;\,{km,nj}}(E_{k,m})
 + R^{q,\lambda,p,-\lambda^{(+)}}_{N;\,{km,nj}}(E_{j,n}) \right)\Big],
\label{eq:ARedfield_counting}
\eea
where $E_{m,n} = E_m - E_n$ and $E_n$ are the energy eigenstates of the system Hamiltonian $\hat H_S$.
At this point, we switched back to the Schr\"{o}dinger representation.
Recall that $p$ and $q$ sum over the different operators that couple to the system.
The other indices, $j,k$, count eigenstates of the system. 
This derivation can be extended to higher orders in the system-bath interaction 
within the projection operator formalism beyond the lowest order Redfield equation. 
We continue and simplify Eq. (\ref{eq:ARedfield_counting})  by performing the secular approximation
so as to decouple population and coherence dynamics.
If the characteristic timescale of the subsystem $\tau_{\textrm{S}} \approx E_{n,m}^{-1}$ is much shorter than the system's relaxation
time $\tau_{\textrm{R}}$,
the function $e^{i\,E_{n,m}t}$ oscillates rapidly over $\tau_{\textrm{R}}$  
and thus the contribution of the coherence terms in the reduced density matrix would be averaged out to zero. 
The resulting population dynamics, $p_{n}(t) \equiv \sigma_{nn}(t)$, is given in Eq. (\ref{eq:population_under_SA_Rs}) in the main text.


\renewcommand{\theequation}{B\arabic{equation}}
\setcounter{equation}{0}  

\section*{Appendix B:  Detailed balance relation with counting parameters}
\label{app:AppB}


To prove the detailed balance relation for a specific bath $\nu$, 
we examine the correlation function $\langle \hat B_\nu^\lambda(s) \hat B_\nu^{\lambda^\prime}(0) \rangle$;
recall that the average is performed with respect to the initial, canonical state
$\rho_\nu 
=e^{-\beta_\nu \hat H_\nu}/Z_\nu$. 
Using the cyclic property of the trace we get
\bea
	\langle \hat B_\nu^\lambda(s) \hat B_\nu^{\lambda^\prime}(0) \rangle = 
		\frac{1}{Z_\nu}\textrm{Tr}_\nu \left[\hat B_\nu^{\lambda^\prime}  e^{i \hat H_\nu (s+i\beta_\nu)}\hat B_\nu^\lambda e^{-i \hat H_\nu (s+i\beta_\nu)} e^{-\beta_\nu\hat H_\nu}  \right].
\eea
From here, we readily organize as the detailed balance relation, 
\bea
	\langle \hat B_\nu^\lambda(s) \hat B_\nu^{\lambda^\prime}(0) \rangle = 
		\langle \hat B_\nu^{\lambda^\prime}(0)  \hat B_\nu^\lambda(s+i\beta_\nu)  \rangle = \langle \hat B_\nu^{\lambda^\prime}(-s-i\beta_\nu)  \hat B_\nu^\lambda(0)  \rangle.
\label{eq:Adetailed_balance}
\eea
Note that Eq. (\ref{eq:Adetailed_balance}) has the counting-field terms switched. 
Moreover,  we can open up the counting-field dependent terms and achieve
\bea
\langle \hat B_\nu^\lambda(s) \hat B_\nu^{\lambda^\prime}(0) \rangle = 
		\frac{1}{Z_\nu}\textrm{Tr}_\nu \left[\hat B_\nu  e^{i \hat H_\nu [s+i\beta_\nu+\frac{1}{2}(\lambda-\lambda^\prime)]}\hat B_\nu e^{-i \hat H_\nu [s+i\beta_\nu+\frac{1}{2}(\lambda-\lambda^\prime)]} e^{-\beta_\nu\hat H_\nu}  \right] 
\nonumber,
\eea
which yields, 
\bea
	\langle \hat B_\nu^\lambda(s) \hat B_\nu^{\lambda^\prime}(0) \rangle &=& 
		\langle \hat B_\nu(0)  \hat B_\nu(s+i\beta_\nu+(\lambda-\lambda^\prime)/2)  \rangle 
\nonumber\\
&=& \langle \hat B_\nu(-s-i\beta_\nu-(\lambda-\lambda^\prime)/2)  \hat B_\nu(0)  \rangle.
\label{eq:detailed_balance_wcounting}
\eea
From here, we immediately deduce that
\bea
	\langle \hat B_\nu^\lambda (s) \hat B_\nu^{-\lambda} (0) \rangle = \langle \hat B_\nu(-s-i\beta_\nu-\lambda)  \hat B_\nu(0)  \rangle,
\eea 
receiving the  detailed balance result,
\bea
	M_{\nu;\,nj,jn}^{\lambda;-\lambda}(\omega) = e^{(\beta_\nu-i\lambda)\omega} M_{\nu;\,nj,jn}(-\omega),
\label{detailed_balance}
\eea
which is essential to much of our analysis.

\renewcommand{\theequation}{C\arabic{equation}}
\setcounter{equation}{0}  

\section*{Appendix C:  Fluctuation relation for two level system under two-terminal transport}
\label{app:AppC}

The steady state fluctuation theorem for entropy production
is a microscopic statement of the second law of thermodynamics \cite{esposito-review,hanggi-review}.
It is crucial to validate it here, so as to establish the thermodynamic consistency of our treatment.
We prove it by verifying the following symmetry of the CGF,
\bea
\mu^\lambda_+ = \mu_+^{-\lambda+i\Delta\beta},
\label{eq:FR}
\eea
with $\Delta\beta=\beta_L-\beta_R$. %
This symmetry, in fact, holds for both eigenvalues in Eq. (\ref{eq:CGFTLS}).
We prove it by showing that
\bea
k_{0\to1}^{\lambda}k_{1\to 0}^{\lambda} = k_{0\to1}^{-\lambda+i\Delta \beta}k_{1\to 0}^{-\lambda+i\Delta \beta},
\label{eq:FRp1}
\eea
for both additive and non-additive interaction models.
First, we recall that for an additive system-bath interaction model, the rate constants satisfy
$k^{\lambda}_{1\to0}=k_{1\to0}^L + k_{1\to0}^{R, \lambda}$,
where 
\bea
k_{0\to 1}^{R, \lambda}&=&\intinf ds \, e^{-i\omega_0s} \langle \hat B_R(s) \hat B_R(0)\rangle \, e^{i\omega_0\lambda},
\nonumber\\
k_{1\to 0}^{R,\lambda}&=&\intinf ds \, e^{i\omega_0s} \langle \hat B_R(s) \hat B_R(0)\rangle \, e^{-i\omega_0\lambda}.
\eea
The $L$-reservoir induced rates take the same form but with $\lambda=0$.
Using the detailed balance relation, $k_{0\to 1}^{\nu}=e^{-\beta_{\nu}\omega_0} k_{1\to0}^{\nu}$,
we can readily show that
\bea
 k_{0\to 1}^{-\lambda+i\Delta \beta} =
\left( k_{1\to 0}^L + k_{1\to0}^{R,\lambda} \right)e^{-\omega_0\beta_L}, \,\,\,\,\
k_{1\to 0}^{-\lambda+i\Delta \beta} =\left( k_{0\to 1}^L+ k_{0\to1}^{R,\lambda} \right)e^{\omega_0\beta_L},
\eea
which proves Eq. (\ref{eq:FRp1}).
It is easy to generalize this proof for systems including more than two states.

In the non-additive case, the rate constants satisfy
\bea
k_{0\to 1}^{\lambda}&=&\intinf d\omega  \, M_L(-\omega_0-\omega)M_R(\omega) \, e^{-i\omega\lambda},
\nonumber\\
k_{1\to 0}^{\lambda}&=& \intinf d\omega  \, M_L(\omega_0-\omega)M_R(\omega) \, e^{-i\omega\lambda}.
\eea
%
Using the detailed balance relation, which holds for each component separately,  
$M_{\nu}(\omega)=e^{\omega\beta_{\nu}} M_{\nu}(-\omega)$,
we confirm that
\bea
k_{0\to 1}^{-\lambda+i\Delta \beta}&=&e^{-\beta_L\omega_0}\intinf d\omega  M_L(\omega_0-\omega)M_R(\omega)e^{-i\omega\lambda}
\nonumber\\
k_{1\to 0}^{-\lambda+i\Delta \beta}&=&e^{\beta_L\omega_0}\intinf d\omega  M_L(-\omega_0-\omega)M_R(\omega)e^{-i\omega\lambda},
\eea
which immediately proves Eq. (\ref{eq:FRp1}). 
A more general proof, for systems with more than two states, was given in Ref. \cite{FCS-strong}.
It is straightforward to generalize these results,  to describe quantum transport in three-terminal setups.
In such models, the system can act as  a quantum absorption refrigerator \cite{QAR},  as discussed in Sec. \ref{sec-engine}.


\end{document}